\documentclass[11pt]{article}
\usepackage[a4paper, margin=1in]{geometry}
\usepackage[numbers]{natbib}
\usepackage{graphicx}
\usepackage{hyperref}
\usepackage{url}
\usepackage{booktabs}
\usepackage{pifont}
\usepackage{float}

\title{LLM-Driven APT Detection for 6G Wireless Networks: A Systematic Review and Taxonomy}

\author{
  Muhammed Golec$^{1,2}$,
  Yaser Khamayseh$^3$,
  Suhib Bani Melhem$^4$,
  Abdulmalik Alwarafy$^5$ \\
  \\
  $^1$School of Electronic Engineering and Computer Science, Queen Mary University of London, UK \\
  $^2$Electrical and Electronics Engineering Department, Bursa Uludag University, Turkey \\
  $^3$The College of Technological Innovation, Zayed University, UAE \\
  $^4$ College of Engineering, Al Ain University, Abu Dhabi, UAE\\
  $^5$ College of Information Technology, UAE University, Al Ain, UAE \\
}

\date{}

\begin{document}
\maketitle

\vspace{1em}
\begin{center}
\textit{This manuscript remains under review.}
\end{center}

\begin{abstract}
Sixth Generation (6G) wireless networks, which are expected to be deployed in the 2030s, have already created great excitement in academia and the private sector with their extremely high communication speed and low latency rates. However, despite the ultra-low latency, high throughput, and AI-assisted orchestration capabilities they promise, they are vulnerable to stealthy and long-term Advanced Persistent Threats (APTs). Large Language Models (LLMs) stand out as an ideal candidate to fill this gap with their high success in semantic reasoning and threat intelligence. In this paper, we present a comprehensive systematic review and taxonomy study for LLM-assisted APT detection in 6G networks. We address five research questions, namely, semantic merging of fragmented logs, encrypted traffic analysis, edge distribution constraints, dataset/modeling techniques, and reproducibility trends, by leveraging most recent studies on the intersection of LLMs, APTs, and 6G wireless networks. We identify open challenges such as explainability gaps, data scarcity, edge hardware limitations, and the need for real-time slicing-aware adaptation by presenting various taxonomies such as granularity, deployment models, and kill chain stages. We then conclude the paper by providing several research gaps in 6G infrastructures for future researchers. To the best of our knowledge, this paper is the first comprehensive systematic review and classification study on LLM-based APT detection in 6G networks. 
\end{abstract}

\section{Introduction}

The rapid development of wireless technologies increases expectations for 6G networks with ultra-low latency and artificial intelligence-based orchestration architecture \cite{ferrag2023edge}. To meet these expectations, 6G networks operate with a heterogeneous architecture where many layers, such as physical and network layers, work together, which means a larger attack surface \cite{osorio2022towards}. The wide attack surface in 6G systems requires that measures be taken against Advanced Persistent Threats (APTs), one of the hidden and long-stage attack methods that are difficult to detect with traditional detection mechanisms \cite{khowaja2024pathway}.

Large Language Models (LLMs) with semantic and contextual reasoning features are one of the most promising developments that can be used against APTs \cite{g2024harnessing}. In particular, it can be used for the detection of APT in 6G networks by analyzing fragmented logs and increasing situational awareness \cite{gulbay2024apt}. Despite this potential, there is no comprehensive taxonomy or systematic analysis in the literature on LLM-based APT detection for 6G networks.

To the best of our knowledge, this paper is the first comprehensive systematic review and classification study on LLM-based APT detection in 6G networks. As a result of the current studies examined by the authors using identification, screening, eligibility, and inclusion (snowballing) techniques, 142 articles were analyzed. Our aimed is to synthesize the intersection of LLM architectures, APT lifecycle modeling, and 6G-specific security challenges and provide insights for future research.

\subsection{Motivation and Contributions}

LLMs and 6G technologies are very recent research areas and their intersection in cyber threat detection, such as APT attacks, has not been sufficiently investigated in the literature. Existing studies are scattered across either LLM-based cybersecurity or 6G network security issues. Furthermore, 6G networks are still in their infancy (expected to become widespread after the 2030s) and contain obstacles for AI and rule-based systems due to issues such as a fragmented structure of source data and end device limitations \cite{rahman2022deep}. For all these reasons, a detailed investigation should be conducted to explore the potential of LLMs in providing explainable detection mechanisms throughout the 6G infrastructure. The main contributions of this paper can be summarized as follows:

\begin{itemize}
\item  We present the first Systematic Literature Review (SLR)-based review focusing on LLM-enabled APT detection in 6G networks. To do this, we searched more than 300 most recent and relevant papers in academic and industrial databases between 2018-2025. As a result of the systematic analysis (Kitchenham's SLR approach and Petersen's Systematic Mapping Study (SMS) \cite{kitchenham2009systematic,petersen2008systematic}), the most relevant 142 papers in the field were obtained.

\item We define five-point research questions to conduct the systematic review (Section 4.2). In line with these questions: (i) Semantic correlation of fragmented logs generated in 6G networks and how it can be used for LLMs threat detection (Section 5.1), (ii) Limitations of 6G encrypted channels and how it can address LLMs visibility and reasoning challenges (Section 5.2), (iii) Challenges of deploying LLM to edge nodes on 6G networks and optimization techniques for these challenges (Section 5.3), (iv) Datasets and modeling techniques used in LLM-based APT detection studies (Section 5.4), and (v) Exploration of publication trends, platform distribution, and reproducibility for LLM-focused APT research (Section 5.5).

\item  LLM deployment models, threat lifecycle stages, optimization strategies for edge inference, and taxonomy studies for dataset types are presented.

\item  Research gaps, such as explainability gaps, dataset scarcity, and 6G orchestration risks, are highlighted through critical analysis. And future directions, such as slice-aware XAI pipelines and unified demand tuning techniques, are highlighted.

\item  A comparison of this paper with 16 previous reviews in the literature. The comparison is made to highlight the novelty and necessity of the paper.

\end{itemize}

\subsection{Article Organization}

\begin{figure}[ht]
	\centering
    \includegraphics[scale=0.5]{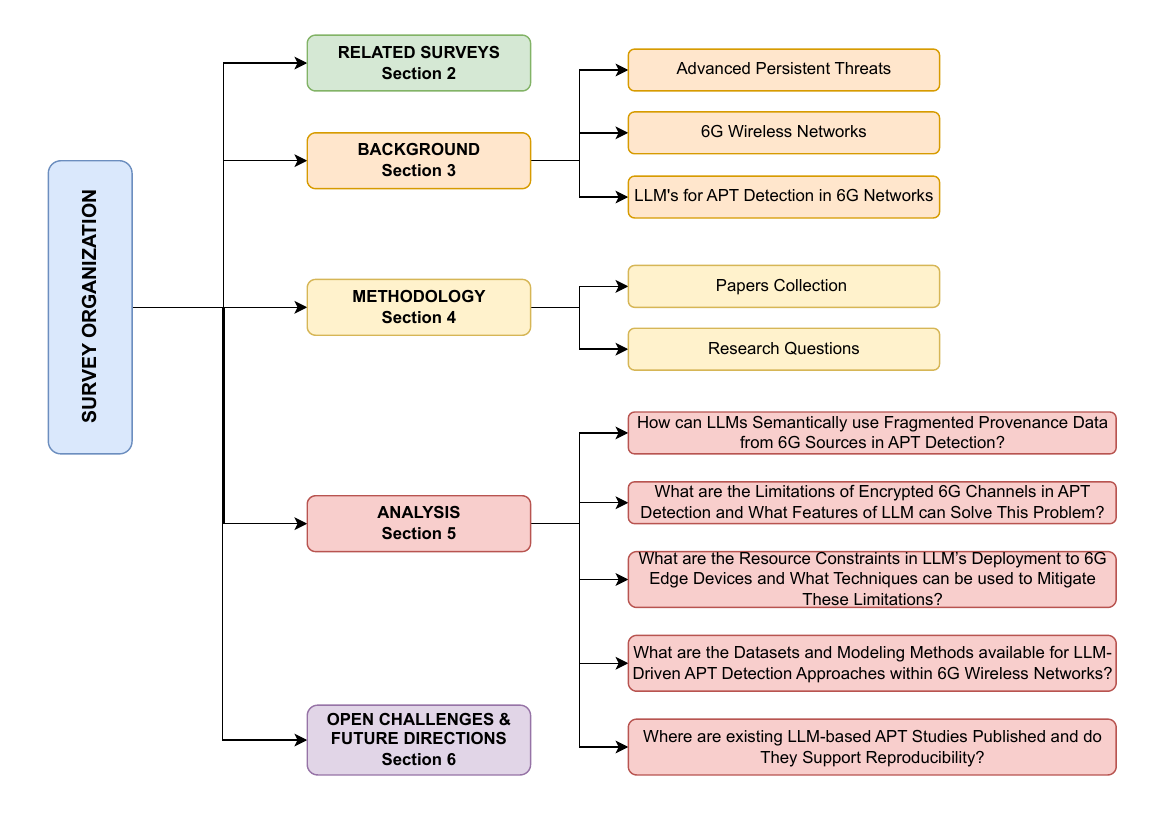}
	\caption{The Organization of the Survey}
	\label{fig:tax}
\end{figure}

Figure \ref{fig:tax}  shows the organizational chart for this paper. Section \ref{sec:related} provides a comparison with related surveys to highlight the novelty of the paper. Section \ref{sec:background} explains the basic background of APTs, 6G networks, and LLMs, and explains their roles in cybersecurity. Section \ref{sec:Methodology} explains the methodological structure, such as the article selection methods and research questions for this systematic review. Section \ref{sec:analysis} provides an in-depth analysis of five key research questions. Section \ref{sec:challengesandfuture} indicates open challenges and future directions for researchers in the relevant research area. Section \ref{sec:conclusions} concludes the paper by summarizing the findings and emphasizing the importance of LLM-based APT detection in 6G.

\section{RELATED SURVEYS}\label{sec:related}

The use of LLMs in the detection and prevention of APTs, which are potential cybersecurity threats in 6G, is still an area that needs to be investigated. The main reason for this is that there are limited datasets in the literature on APTs and 6Gs can only be modeled simulation-based. When the literature is examined, it is seen that although there are surveys focused on 6G, APT, LLMs, and LLM-based security, however, there is no systematic review and taxonomy study that addresses LLM, APT, and 6G in a combined manner.

\textbf{LLM-Focused Cybersecurity Surveys}: Several recent studies in the literature have examined the role of LLMs in the field of cybersecurity. Hassanin et al. \cite{1} provide an overview of the role of LLMs in applications such as threat intelligence and phishing detection in their review. In \cite{2,3,4,7}, they examine the architectures used for LLM-based attack detection and threat analytics in a more systematic way. Zuo et al. \cite{14} presented an analysis study examining LLM's usage for APTs. This study investigated the semantic augmentation of LLM's (such as GPT-4o) origin logs for APT detection. However, this study is superficial, not a survey or taxonomy, and does not include the 6G context. In another study, the authors present a review of language models (including APT), but do not include any information about 6G and do not use a formal methodology such as SLR \cite{9}. Although LLM's sheds light on the applications in cybersecurity, none of these studies cover 6G and its limitations and opportunities.

\textbf{APT-Focused Surveys}: Some of the literature studies investigated APT detection using DL and rule-based learning. In \cite{6,8}, classifications and threat lifecycle analyses of APTs are examined, while in \cite{15} DL-based cyber attack detection systems (partially addressing APT) are investigated. Although all these survey studies partially or in detail mention APTs, none of them provide detailed information about LLM-based approaches or 6Gs (such as network layer dynamics).

\textbf{6G Focused Surveys}: Another area of survey research in the literature examines the technical foundations of 6G. Shen et al. \cite{10} covers five main aspects of 6G (such as spectrum and positioning) in detail. In \cite{5, 11, 13, 16}, important components in 6G (such as IoT integration and federated learning) are comprehensively examined. In another survey study, Sun et al. \cite{12} investigate the importance and use of explainable AI (XAI) in 6G network slicing and vehicle contexts. However, none of these studies address LLM and security issues.

Among the reviewed literature studies, \cite{2, 3, 4, 7, 15} examine model architectures and use cases in detail in their focused research topic using systematic research methodology such as PRISMA. Furthermore, some of the studies \cite{3,7} provide binary taxonomies of cybersecurity tasks, while in \cite{5, 10, 11, 16} they strongly address 6G at the architecture and protocol level. However, none of the reviewed articles address LLM, APT, and 6G in a unified manner.

\subsection{Critical Analysis}

Table \ref{tab:relatedworks} provides a comparison of 16 recent survey studies with this paper. When the table is examined, it is seen that this paper fills the following three critical gaps:

\begin{itemize}
    
\item  \textbf{Combining consideration of LLM, APT, and 6G}: None of the reviewed studies simultaneously address the intersection of LLM-based threat detection, APT lifecycle modeling, and 6G network features.

\item  \textbf{Providing a detailed taxonomy for APT detection}: The vast majority of studies are in the form of a general survey, and those that do include a taxonomy lack details such as the lifecycle of APTs.

\item  \textbf{Providing a comparative synthesis across fields}: Few of the reviewed studies include multidimensional comparisons (such as methodology, model types). The lack of such comparative syntheses makes it difficult to assess the overlap and gaps between the topics covered by the survey.

\end{itemize}

\begin{table*}[ht]
\caption{Comparison of Our Systematic Review and Taxonomy with Existing Survey Studies}
\label{tab:relatedworks}
\resizebox{\textwidth}{!}{%
\begin{tabular}{@{}ccccccccc@{}}
\toprule
\textbf{Paper} & \textbf{Focus Area} & \textbf{6G-Specific} & \textbf{APT-Specific} & \textbf{LLM-Specific} & \textbf{Type} & \textbf{SLR Methodology} & \textbf{Publisher} & \textbf{Year} \\ \midrule
\cite{1} & General Cyber Defence & \ding{55} & \ding{55} & \ding{51} & Review & \ding{55} & Arxiv & 2024 \\
\cite{2}& LLMs in Cybersecurity & \ding{55} & \ding{51} & \ding{51} & Systematic Review & \ding{51} & Arxiv & 2024 \\
\cite{3}& IDS with Transformers \& LLMs & \ding{55} & \ding{51} & \ding{51} & Review and Taxonomy & \ding{51} & Elsevier & 2024 \\
\cite{4}& LLMs in Cybersecurity & \ding{51} & \ding{51} & \ding{51} & Systematic Survey & \ding{51} & IEEE & 2025 \\
\cite{7}& LLMs in Cyber Threat Detection & \ding{55} & \ding{51} & \ding{51} & Systematic Review & \ding{51} & Elsevier & 2024 \\
\cite{14} & LLM-Augmented Provenance for APT Detection & \ding{55} & \ding{51} & \ding{51} & Research Paper & \ding{55} & Sandia TR & 2025\\
\cite{9}& PLMs/LLMs in Cybersecurity & \ding{55} & \ding{51} & \ding{51} & Review & \ding{55} & IEEE (ISDFS) & 2024 \\
\cite{6}& APT Analysis \& Countermeasures & \ding{55} & \ding{51} & \ding{55} & Review and Taxonomy & \ding{55} & Springer & 2019\\
\cite{8}& APT Detection Techniques & \ding{55} & \ding{51} & \ding{55} & Survey & \ding{51} & TechScience (CMC) & 2024 \\
\cite{15} & DL Techniques for IDS & \ding{55} & \ding{51} & \ding{55} & Systematic Survey & \ding{51} & ACM & 2025 \\
\cite{10} & 6G Architectures \& Networking & \ding{51} & \ding{55} & \ding{55} & Survey & \ding{55} & ACM CSUR & 2023 \\
\cite{5}& Federated Learning in 5G/6G Cybersecurity & \ding{51} & \ding{55} & \ding{55} & Comprehensive Survey & \ding{51} & IEEE & 2025 \\
\cite{11} & 6G and IoT Integration & \ding{51} & \ding{55} & \ding{55} & Comprehensive Survey & \ding{55} & IEEE & 2022 \\
\cite{13} & Optimization \& Performance of LIS in 6G & \ding{51} & \ding{55} & \ding{55} & Survey & \ding{55} & IEEE Access & 2020 \\
\cite{16} & 6G Technologies \& Architectures & \ding{51} & \ding{55} & \ding{55} & Survey & \ding{55} & IEEE & 2022 \\
\cite{12} & Explainable AI for 6G & \ding{51} & \ding{55} & \ding{55} & Systematic Survey & \ding{51} & IEEE OJ-COMS & 2025 \\
\textbf{Our Paper} & \textbf{LLMs for APT Detection in 6G} & \ding{51} & \ding{51} & \ding{51} & \textbf{Systematic Review and Taxonomy} & \ding{51} & \textbf{Computer Science Review (Planned)} & \textbf{2025} \\ \bottomrule
\end{tabular}%
}
\end{table*}

\section{BACKGROUND}\label{sec:background}

This section provides the basic background necessary for the reader to better understand the concepts related to LLM-based APT detection in 6G networks.

\subsection{Advanced Persistent Threats}

Advanced Persistent Threats (APT) are one of the most effective cyber attacks known due to their characteristics, such as stealth and longevity. This subsection defines APT and explains its key characteristics, lifecycle, and attacker behaviors (TTPs). Then, a comparison of traditional attacks and APTs is provided for APT.

\subsubsection{Key Characteristics of APT}

APTs use multiple vectors to gain long-term access to an IT environment and are like an attacker with significant expertise and resources \cite{ross2020enhanced}. They have three basic characteristics \cite{xing2020review}:

\begin{itemize}
    
\item  \textbf{Advanced}: These attacks specialize in zero-day attacks and tactics to evade detection.

\item  \textbf{Persistent}: They encourage new APT attacks by leaving backdoors in the systems they penetrate.

\item  \textbf{Threat}: They carry out attacks such as espionage, sabotage, or exfiltration of critical data from the systems.

\end{itemize}

\subsubsection{The Lifecycle of APTs}

\begin{figure}[ht]
\centering
\includegraphics[width=0.5\textwidth]{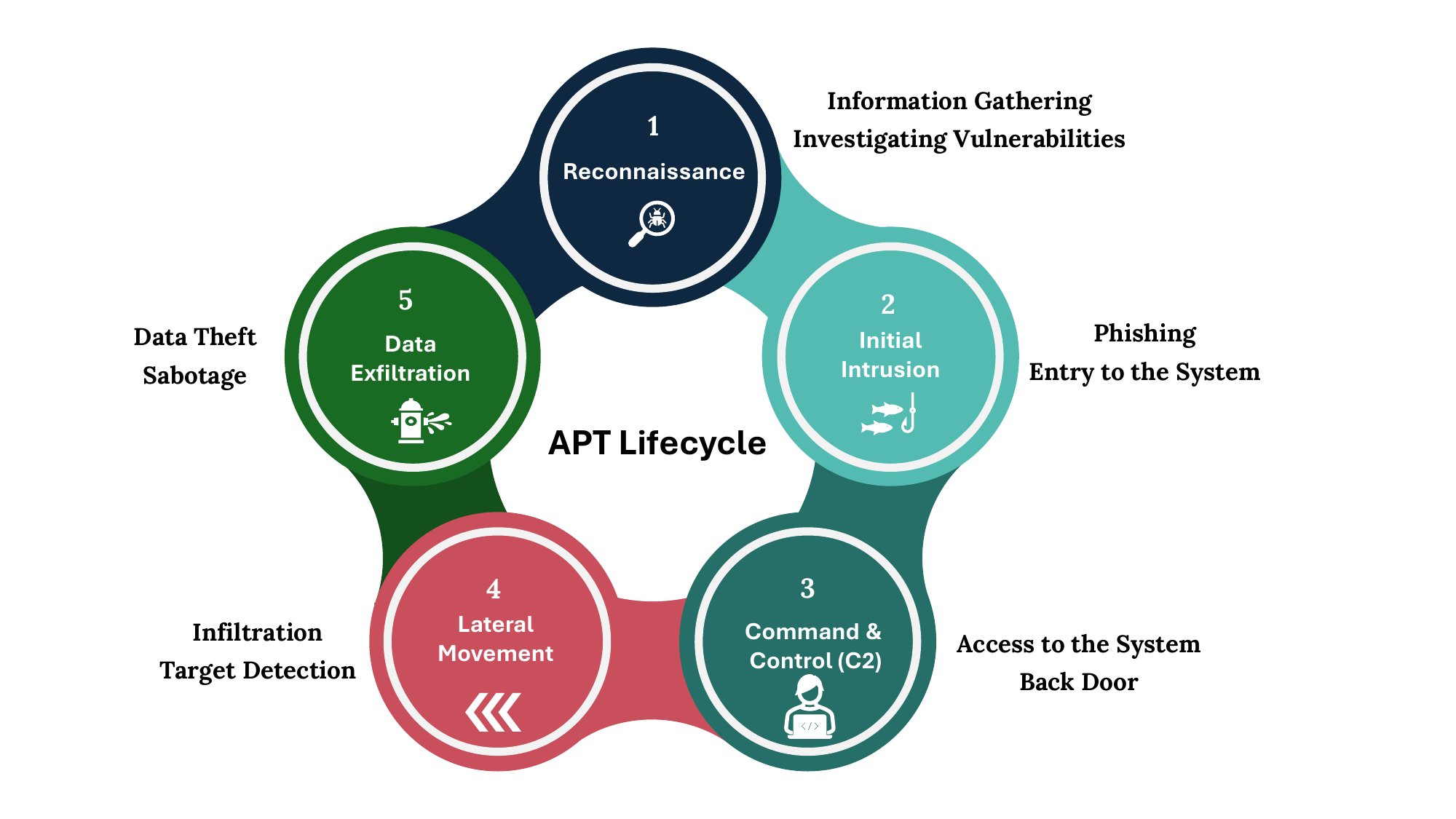}
\caption{The five-stage Lifecycle of an APT}
\label{fig:apt_lifecycle}
\end{figure}

The lifecycle for APT attacks is shown in figure \ref{fig:apt_lifecycle} and can be summarized in five basic stages \cite{6,8,9}:

\begin{itemize}

\item \textbf{Reconnaissance}: This is the first stage of the attack, and information about the target is collected (Open Source Scanning (OSINT)), and system vulnerabilities and weaknesses are investigated.

\item \textbf{Initial Intrusion}: The entry into the target system is achieved through methods such as phishing and malware.

\item \textbf{Command and Control (C2)}: Preparation of the infrastructure to communicate with the APT inserted into the target system (such as backdoor channels).

\item \textbf{Lateral Movement}: Infiltration of other devices connected to the same network within the system and detection of high-value targets.

\item \textbf{Data Exfiltration}: The final stage involves malicious operations such as exfiltration of data in the target system using APT and system sabotage.

\end{itemize}

\subsubsection{Tactics, Techniques, and Procedures (TTPs)}

\begin{figure}[ht]
\centering
\includegraphics[width=0.5\textwidth]{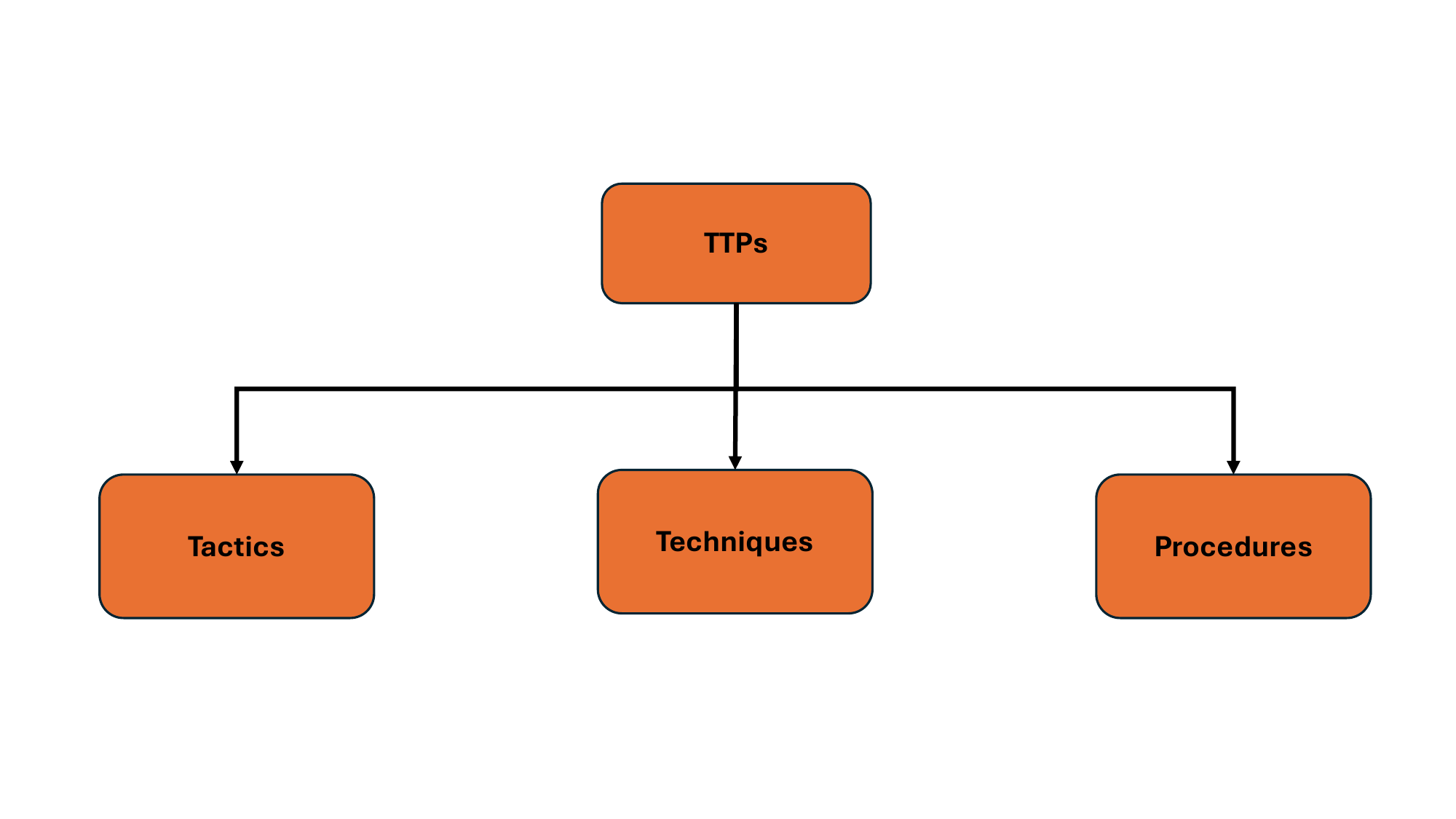}
\caption{The Hierarchical Structure of TTPs}
\label{fig:ttp_structure}
\end{figure}

TTPs are shown in Figure \ref{fig:ttp_structure} and are the framework used to classify the behavior of an APT attack. It can be defined as follows \cite{arulkumar2023apt}:

\begin{itemize}
    
\item  \textbf{Tactics}: Used to define the goal of the attack, such as gaining access to a system.

\item  \textbf{Techniques}: Refers to the technique used to achieve this goal. An example would be DLL injection.

\item  \textbf{Procedures}: Refers to the way the attack is implemented, such as sending a special email.

\end{itemize}

\subsubsection{APT vs. Traditional Attacks}

APTs and traditional attacks differ from each other in many ways, such as target, tactics, duration, and these differences are shown in Table \ref{tab:llm3}. Traditional attacks aim to cause general damage and aim for quick gain, while APTs are long-term and professional attacks (usually state-sponsored) \cite{xing2020review}. Traditional attacks identify weak systems by simultaneously attacking many targets, while APTs are more target-oriented and the attack process is carried out in a long and sneaky way \cite{lv2022review}.

\begin{table}[ht]
\centering
\caption{Comparative Characteristics: Traditional Attacks vs. Advanced Persistent Threats (APT)}
\label{tab:llm3}
\resizebox{\linewidth}{!}{%
\begin{tabular}{@{}lll@{}}
\toprule
\textbf{Attribute} & \textbf{Traditional Attack} & \textbf{Advanced Persistent Threat (APT)} \\ \midrule
Target             & Broad or Random            & Highly Specific                           \\
Duration           & Short-lived                & Prolonged (Months or Years)               \\
Entry Vector       & Known Exploits             & Custom Zero-Days, Spear Phishing          \\
Goal               & Financial Gain, Disruption & Espionage, Strategic Access               \\
Tools Used         & Commodity Malware          & Tailored, Multi-Stage Toolkits            \\ \bottomrule
\end{tabular}%
}
\end{table}

\subsection{6G Wireless Networks}

\subsubsection{Architectural Foundations of 6G}

6G is the new generation of wireless communication paradigm that emerges with the integration of advanced physical technologies and software-defined network solutions \cite{ullah2024survey}. In order to provide uninterrupted communication in the 6G architecture, it is a heterogeneous structure (Ultra-Dense Heterogeneous Networks) that combines three basic layers: terrestrial, aerial, and satellite \cite{song2025empirical}. In order to reach a data rate of more than 1 Tbps, technologies such as Terahertz (THz) communication and Visible Light Communication (VLC) are used \cite{ariyanti2020visible}. In addition, power-sensitive technologies such as Reconfigurable Smart Surfaces (RIS) and Software-Defined Metasurfaces (SDM) are used to reduce latency \cite{elamassie2023free}.

\begin{figure}[ht]
    \centering
    \includegraphics[width=0.5\textwidth]{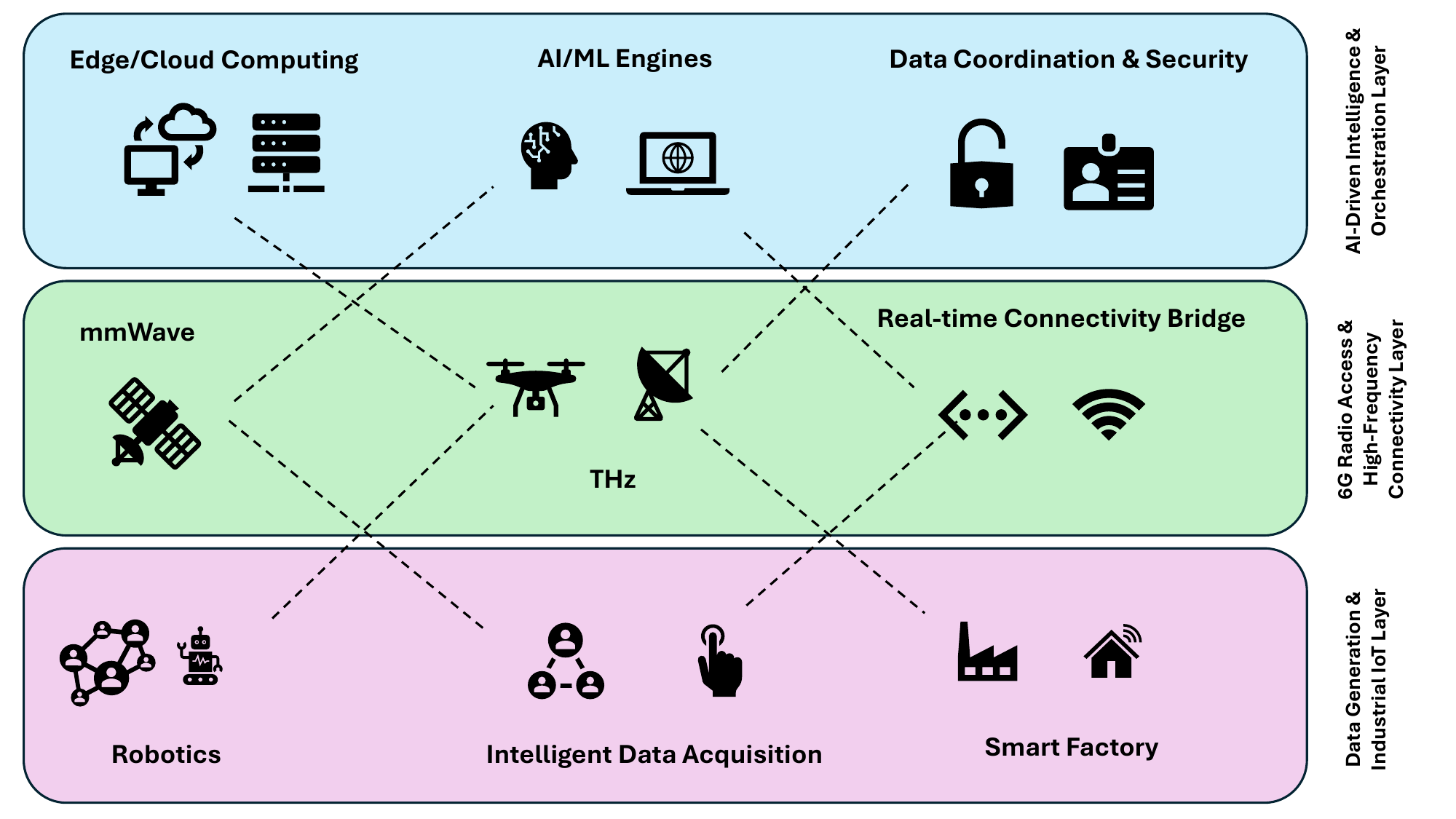}
    \caption{6G Architectural Pillars and Deployment Layers}
    \label{fig:6g_architecture}
\end{figure}

The 6G networks with heterogeneous architecture shown in Figure \ref{fig:6g_architecture} try to reduce computational loads with edge devices and AI-based systems \cite{mao2023security}. While 5G architectures use a centralized system, in the 6G architecture, thanks to decentralization, network slices can be optimized autonomously via AI-based engines. However, despite these advantages, the heterogeneous and decentralized architecture offers a large attack surface and is exposed to cyberattack threats \cite{mao2023security}.

\subsubsection{Key Features of 6G}

Table \ref{tab:5g_vs_6g} shows a feature comparison of 5G and 6G, and as can be seen, 6G (key features) is superior in every aspect. 6G is expected to work in harmony with real-time holography and trusted autonomous systems once it is available for daily use \cite{bhide2024review}. 6G relies on AI-powered protocols and advanced infrastructure capabilities to meet these demands \cite{singh2025ai}.

\begin{table}[ht]
\centering
\caption{Comparison of Key Features Between 5G and 6G Wireless Networks}
\label{tab:5g_vs_6g}
\resizebox{\linewidth}{!}{%
\begin{tabular}{|l|l|l|}
\hline
\textbf{Feature} & \textbf{5G} & \textbf{6G} \\
\hline
Latency & Approximately 1 millisecond & Less than or equal to 0.1 milliseconds \\
\hline
Data Rate & Up to 20 Gbps & At least 1 Tbps \\
\hline
Frequency Range & Sub-6 GHz and millimeter wave & Sub-Terahertz (Sub-THz) and Visible Light Communication (VLC) \\
\hline
Architecture & Centralized network control & Distributed and AI-powered network architecture \\
\hline
Security & Add-on security features & Built-in, intent-aware security mechanisms \\
\hline
\end{tabular}
}
\end{table}

\subsubsection{Vulnerabilities and Threat Landscape in 6G}

Despite the high speed and wide infrastructure opportunities they offer, 6G networks also carry risks such as misconfiguration and hostile exploitation due to AI-based control logic and network software such as SDN/NFV \cite{abdulqadder2022sliceblock}. If vertical slicing and segmentation operations in networks do not work correctly, they become vulnerable to lateral attacks (sourced by APT's etc.) \cite{siriwardhana2021ai}.

\begin{figure}[ht]
    \centering
    \includegraphics[width=0.5\textwidth]{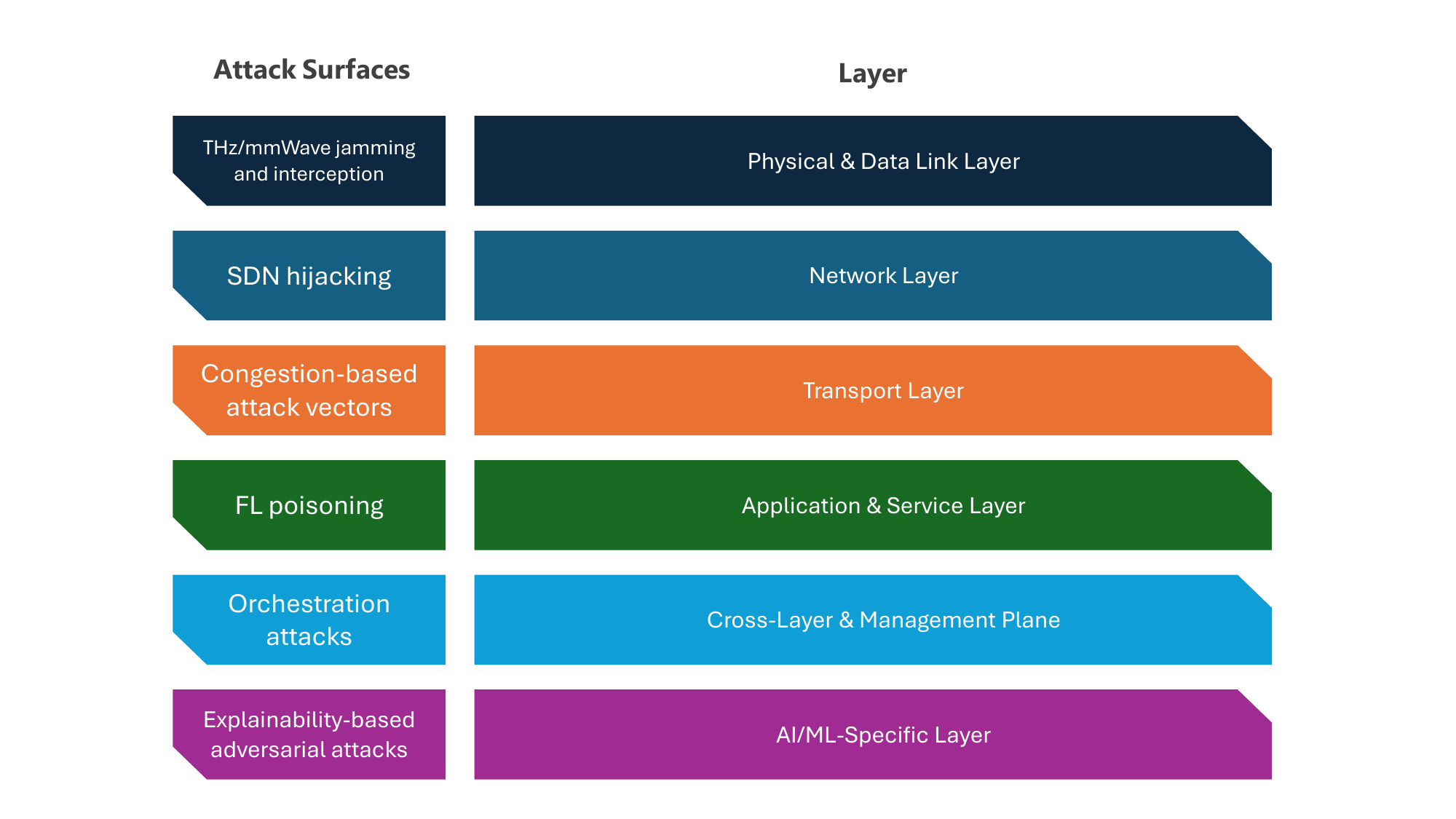}
    \caption{Illustration of Potential Attack Surfaces Across the Hierarchical 6G Network}
    \label{fig:6g_attack_surfaces}
\end{figure}

Possible potential attacks that may occur in 6G are shown in Figure \ref{fig:6g_attack_surfaces}. Attack types can range from physical layer compression to manipulation. Another potential danger is that the AI mechanisms responsible for 6G orchestration are vulnerable to attack and data leakage in cases where RIS and THz communication channels are not properly set \cite{naeem2023security}

\subsubsection{6G-Specific Challenges for APT Detection}

APTs are expected to threaten the rapid lateral movements that 6G will bring to our lives \cite{suomalainen2025cybersecurity}. In addition, behavioral detection becomes more complex for traditional signature-based intrusion detection systems (IDS) due to architectural features such as encrypted layers and dynamic topologies \cite{smiliotopoulos2024detecting}.

Another challenge for 6G networks is the scarcity of APT datasets, which makes it difficult to train security models \cite{mishra2024harnessing}. Another challenge is the fragmented nature of source logs, as this limits the correlation between layers that can be used in APT detection \cite{milajerdi2019holmes}.

\subsubsection{Research Trends Integrating 6G and AI for Security}

Literature studies investigate the use of FL at edges to detect attacks while also concerning privacy \cite{yang2022security}. In addition, XAI methods for decision-making mechanisms are another frequently investigated method \cite{bertrand2022cognitive}. Beyond these, mapping TTPs and analyzing logs with LLM's based systems are promising \cite{zhang2024unittp}. However, since edge devices are resource-constrained and storage-limited devices, these limitations should be taken into consideration when deploying LLM's at edges, and strategies such as model distillation should be applied \cite{agrawal2025efficient}.

\subsection{LLM's for APT Detection in 6G Networks}
\subsubsection{Overview of LLM Architectures and Security-Oriented Specializations}

Developed on Transformer architecture, LLM's have made a great breakthrough in the field of AI, and these models provide representation learning by making sense of the context \cite{wu2024llm}. In other words, these LLM's indicate the ability to understand the meanings of words in context beyond their dictionary meanings. Thanks to this ability, they achieve great success in natural language understanding (NLU) and generation (NLG) tasks \cite{karanikolas2023large}.

\begin{figure}[ht]
    \centering
    \includegraphics[width=0.5\textwidth]{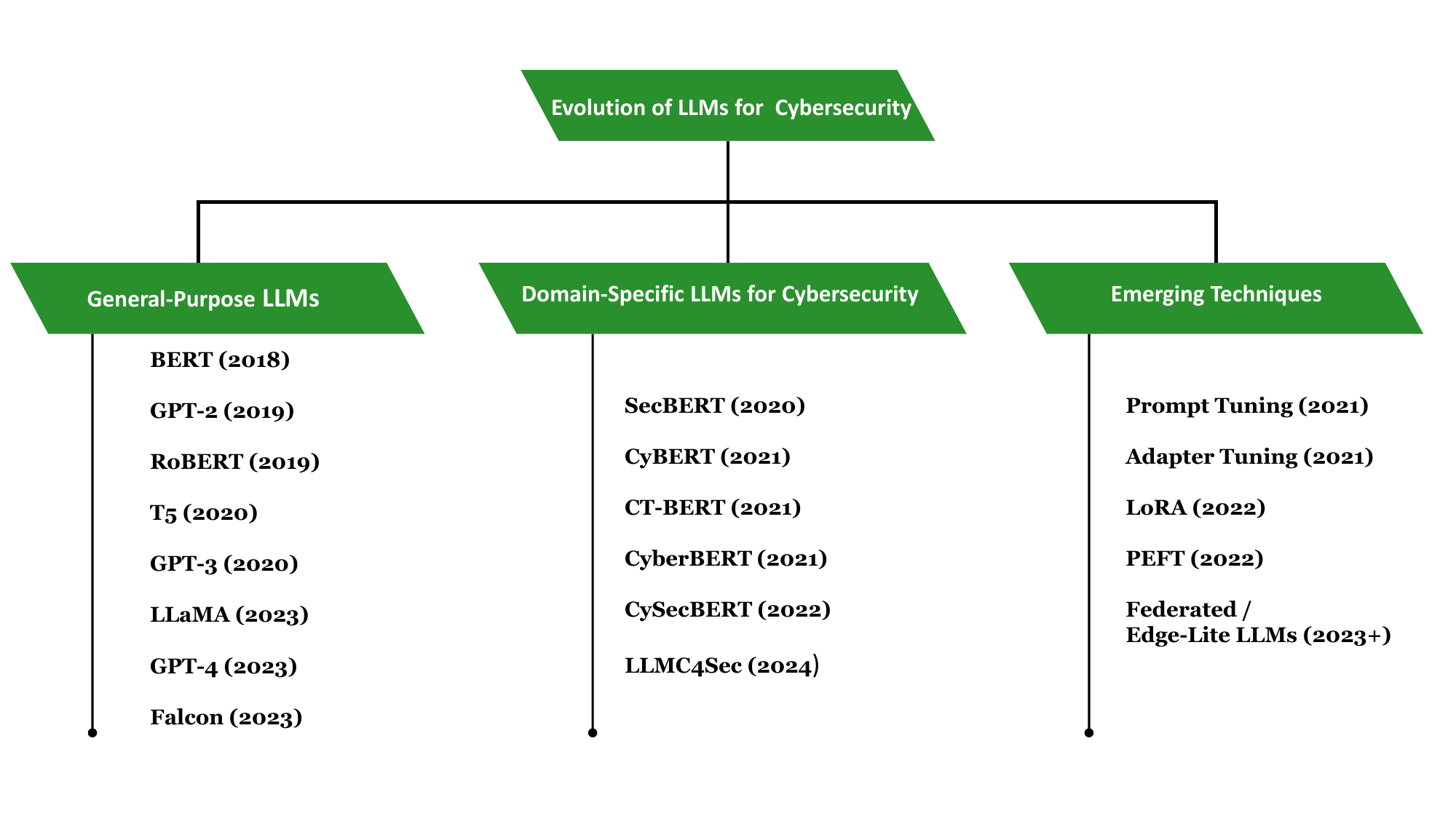} 
    \caption{Evolution of Large Language Models (LLMs) for Cybersecurity}
    \label{fig:llm_evolution}
\end{figure}

This technological development (LLMs) has begun to be used in many areas, especially in cybersecurity, and the evolution of LLMs in cybersecurity use is shown in Figure \ref{fig:llm_evolution}. These areas of use can be examined under three main headings \cite{di2024use,padiu2024extent}:

\begin{itemize}
    
\item  \textbf{General-Purpose}: LLM models that can be trained with large text collections and used for various purposes.

\item  \textbf{Domain-Specific}: LLM models trained (fine-tuned) using purpose-oriented cybersecurity data.

\item  \textbf{Emerging Techniques}: These are methods that make LLM's models lightweight and specific to their intended use.

\end{itemize}

General-purpose models such as BERT (2018), GPT-2 (2019) have shown success in text classification and question-answer tasks, and with the customization of these models, domain-specific models such as SecBERT (2020), CyBERT (2021) have been developed and started to be used in special tasks such as malware detection (such as APT). And studies on emerging techniques continue to increase the performance of these models.

\subsubsection{Applications in Cybersecurity and APT Detection}

\begin{figure}[ht]
    \centering
    \includegraphics[width=0.5\textwidth]{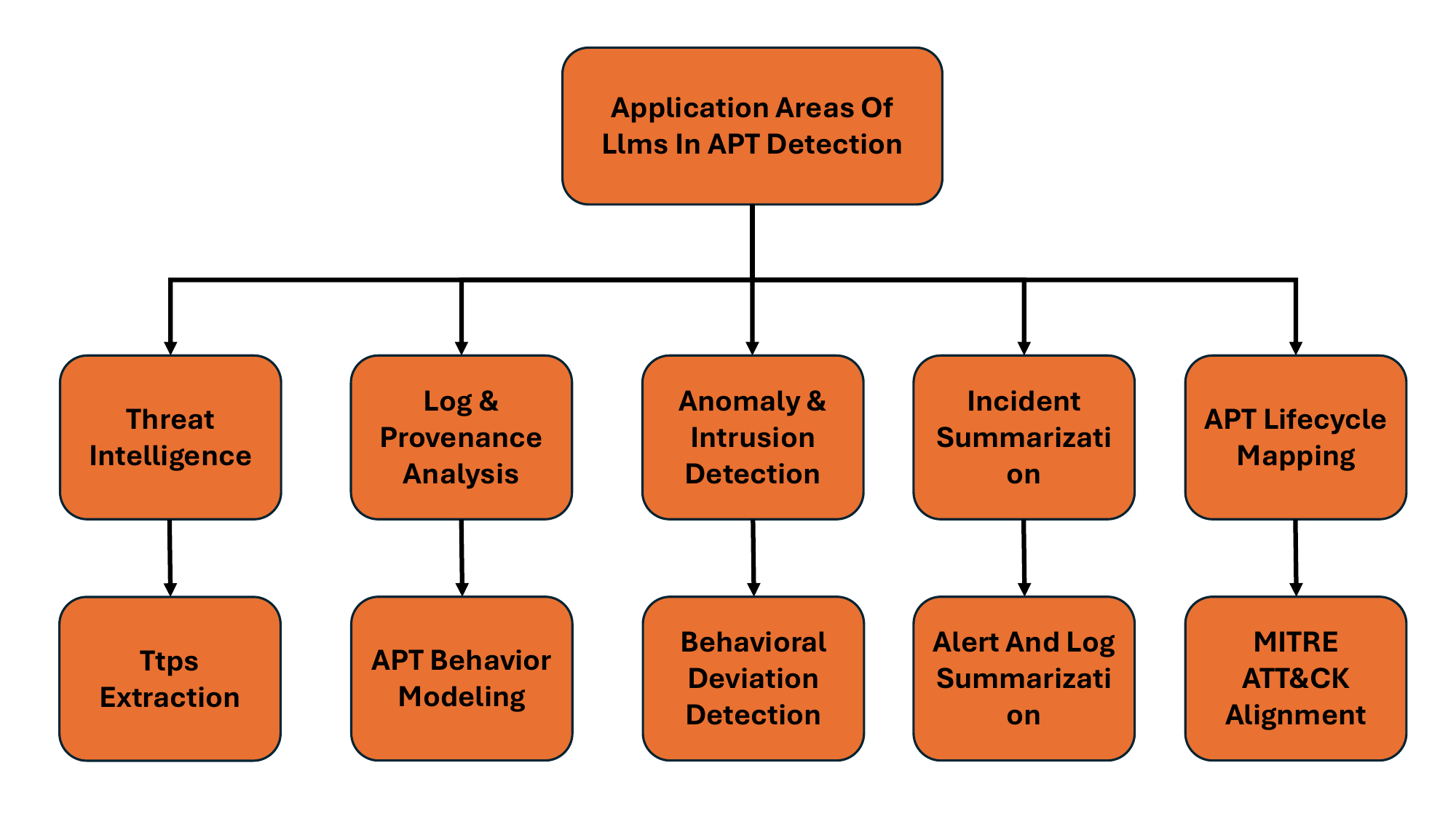}
    \caption{Application Areas of LLMs in APT Detection}
    \label{fig:llm_apt_areas}
\end{figure}

LLMs have been used in cybersecurity for multi-APT detection and response, returning based on attack type \cite{chen2024survey}. LLM's features, such as contextual reasoning and linguistic understanding, make it particularly suitable for APTs with multi-stage attacks \cite{shenoy2024extended}. Figure \ref{fig:llm_apt_areas} shows a tree structure for LLM application areas in APT detection. As can be seen from the figure, LLMs are versatile in the cybersecurity context \cite{hassan2020tactical,liu2022rapid,buchta2024advanced}:

\begin{itemize}

\item  \textbf{Threat Intelligence}: LLM's models can extract TTPs using open-source data such as threat reports.

\item  \textbf{APT Behavior Modeling}: Logs and lineage data can be used to semantically interpret multi-stage APTs.

\item  \textbf{Anomaly Detection}: LLMs context-aware feature can detect anomalous behavior (network and system logs).

\item  \textbf{Alert Triage and Incident Response}: Natural language summarization translates alerts into insights to extract meaningful information (helpful for analysts).

\item  \textbf{TTP Alignment}: LLMs fine-tuning can map hostile behaviors for low-fire systems to MITRE ATT\&CK stages.

\end{itemize}

In dynamic networks (especially 6G networks), the GPT family provides situational awareness in complex architectures where traditional detection methods can struggle.

\subsubsection{LLM Integration Challenges in 6G Edge Environments}

\begin{figure}[ht]
    \centering
    \includegraphics[width=0.5\textwidth]{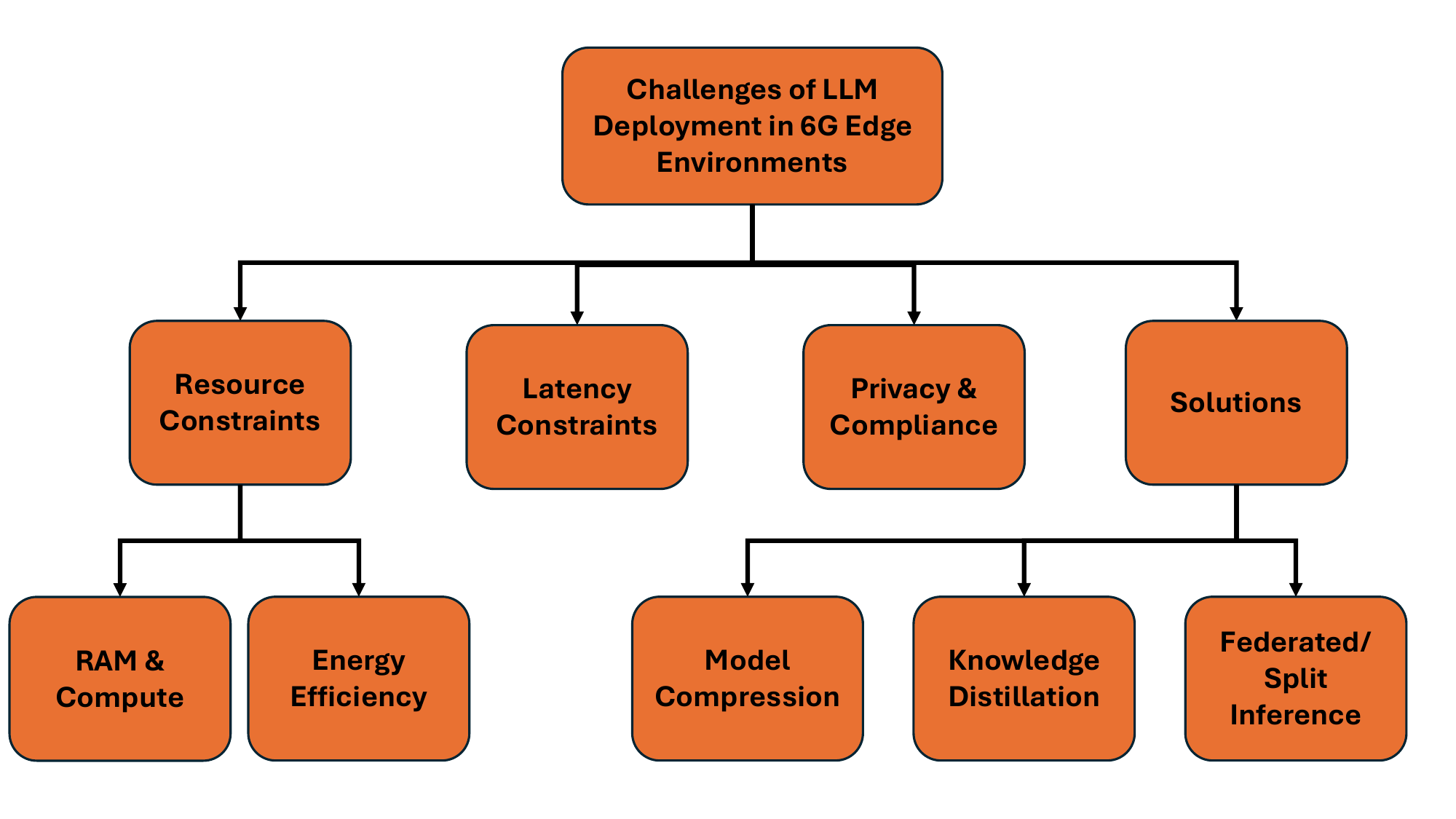}
    \caption{Challenges of LLM Deployment in 6G Edge Environments}
    \label{fig:llm_6g_challenges}
\end{figure}

Edge devices positioned close to the data source have advantages such as low latency and low bandwidth usage, but also disadvantages such as heterogeneous structure and limited processing power \cite{golec2020biosec}. For this reason, problems arise due to these limitations when LLMs are deployed on 6G-based edge devices. Figure \ref{fig:llm_6g_challenges} summarizes these limitations and possible solutions \cite{ferrag2023edge, osorio2022towards}:

\begin{itemize}
    
\item \textbf{Resource Constraints}: Since LLMs require high processing power and storage, their implementation on resource-constrained 6G edge devices (RAM \& Compute Energy Efficiency) is one of the major problems that can be encountered.

\item \textbf{Latency Constraints}: Edge devices, which are expected to offer a low latency advantage due to being positioned close to the data source, may lose this advantage due to the high computational time of LLMs.

\item \textbf{Privacy \& Compliance}: Sensitive data such as biometrics must take into account some privacy concerns when processed on edge devices \cite{golec2020biosec, golec2023healthfaas}.

\end{itemize}

The methods to solve these challenges can be summarized as follows \cite{hafi2024split, tao2024federated}:

\begin{itemize}

\item  \textbf{Compression}: LLMs can be downgraded to lower versions to reduce memory and processing load.

\item  \textbf{Knowledge Distillation}: Information obtained from large models can be transferred to smaller models to minimize performance loss.

\item  \textbf{Federated/Split Inference}: Both privacy and efficiency can be increased by distributing the components of the LLM model to different edge nodes and processing them.
 
\end{itemize}

\subsubsection{APT Detection-Specific Benefits of LLMs in 6G Context }

LLM models have great potential in APT threat reduction studies in 6G networks, which are expected to be used in the near future. This potential stems from the success of LLM models in establishing semantic correlations between data types and their ability to analyze the attack lifecycle as a whole \cite{yao2024survey}. The contributions of LLM models for APT detection in 6G networks can be generalized as follows \cite{wang2025large,friha2024llm}:

\begin{itemize}
    
\item  \textbf{Cross-Layer Fusion}: It can be used in the detection of multi-vector attacks by combining log records from the control and user planes and the cloud layers.

\item  \textbf{Lifecycle Prediction}: LLM models can use past attack data to predict the next step in the APT kill-chain.

\item  \textbf{Semantic Generalization}: LLM models can capture attacks in a contextual manner in encrypted and hidden attack situations where traditional systems are inadequate.

\end{itemize}

\begin{figure}[ht]
    \centering
    \includegraphics[width=0.5\textwidth]{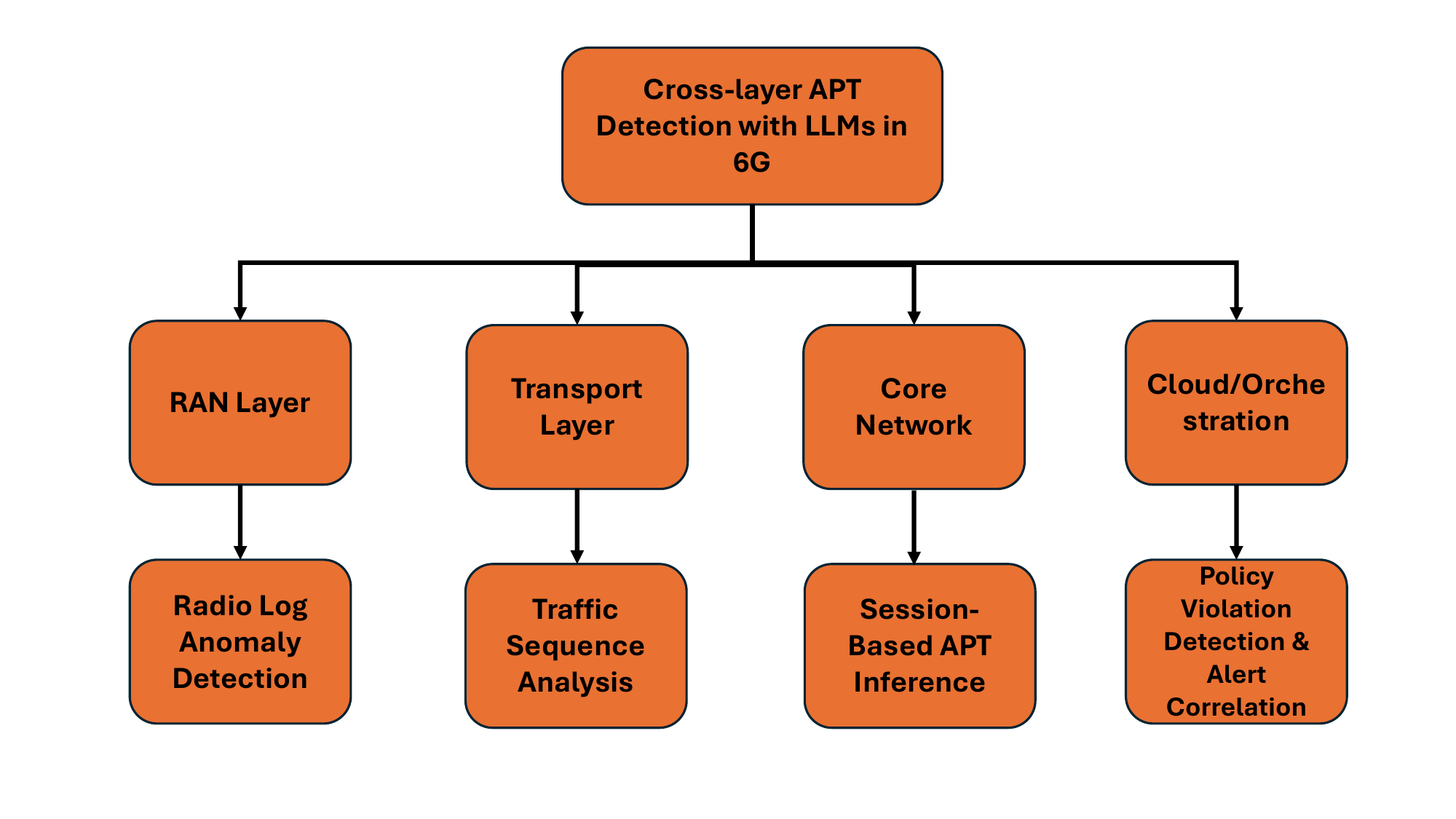}
    \caption{Cross-layer APT Detection with LLMs in 6G}
    \label{fig:llm_crosslayer_6g}
\end{figure}

Figure~\ref{fig:llm_crosslayer_6g} shows the contributions of LLM models for layers. As can be seen from the figure, LLM models can perform multi-layered threat modeling by performing detection not only at the packet level but also at various levels of the network.

\begin{figure}[ht]
    \centering
\includegraphics[width=0.5\textwidth]{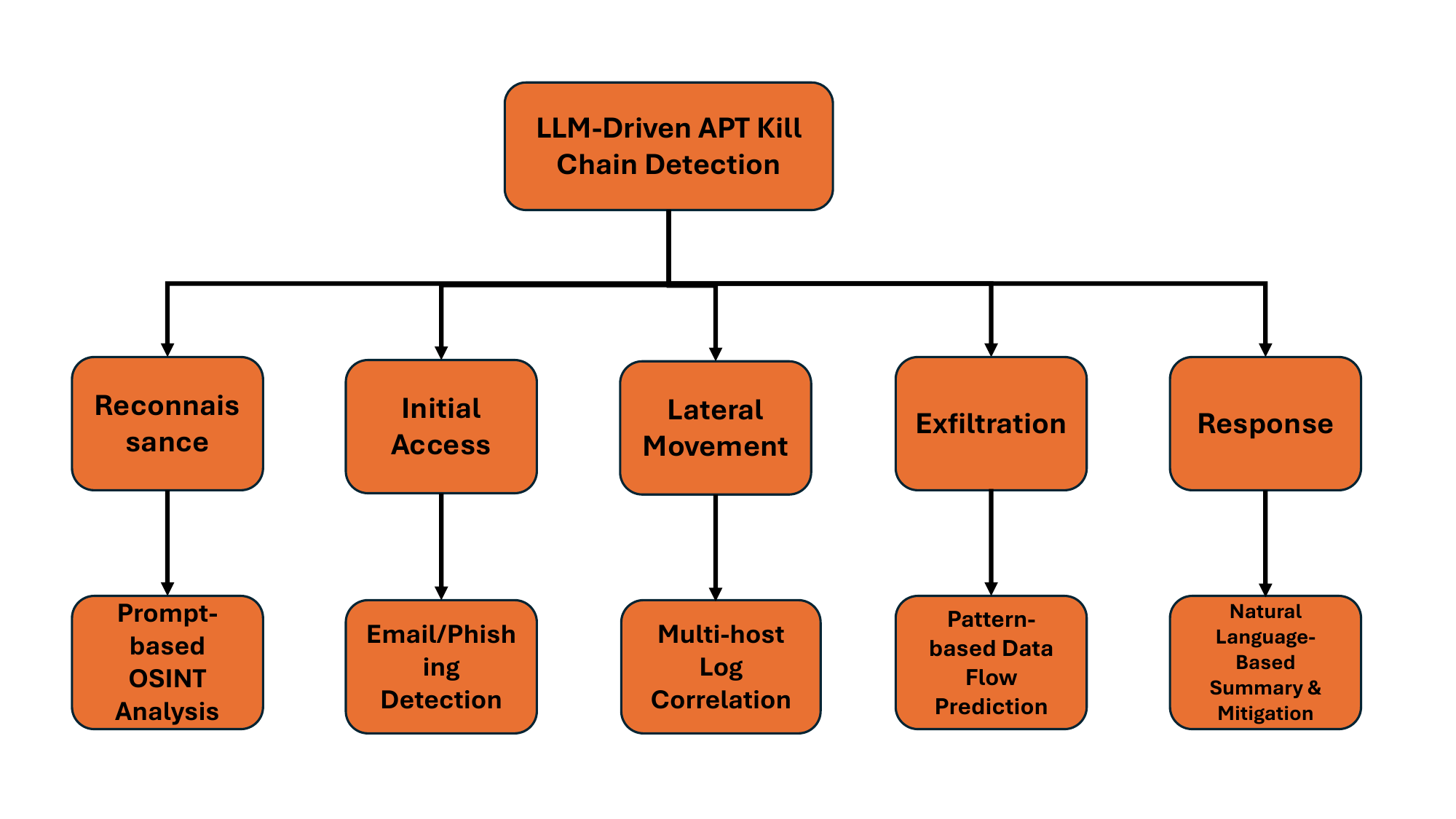}
    \caption{LLM-Driven APT Kill Chain Detection}
    \label{fig:llm_killchain}
\end{figure}

In addition, LLM models not only interpret the behavior of APT attacks but also provide insight into the attack lifecycle phases and response mechanisms. Figure \ref{fig:llm_killchain} shows the tasks that LLMs undertake in the APT kill chain model.

\subsubsection{Proposed Taxonomy: LLM-Driven APT Detection in 6G}

\begin{figure}[ht]
    \centering
    \includegraphics[width=0.5\textwidth]{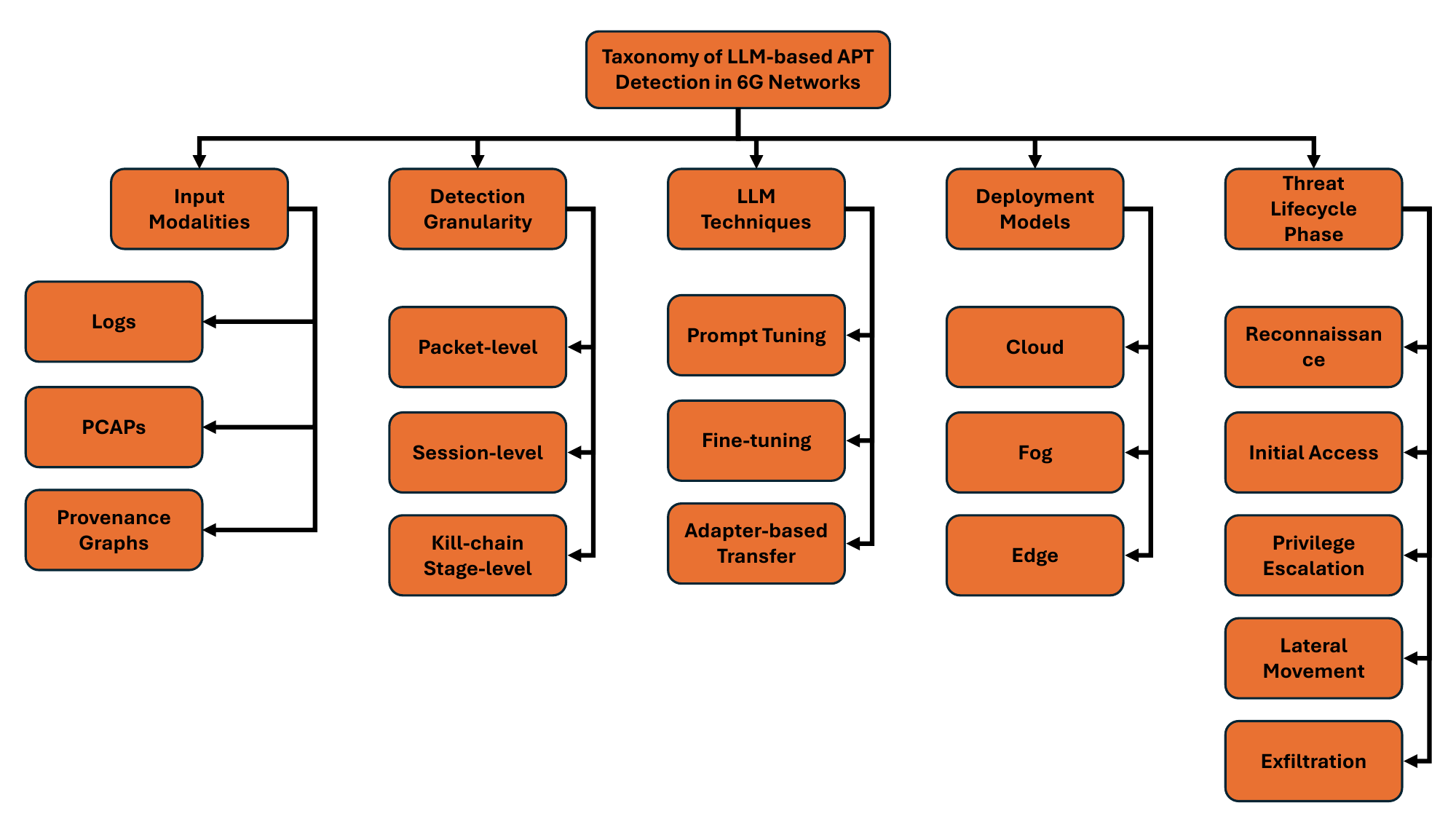}
    \caption{LLM-Based APT Detection Taxonomy in 6G Networks}
    \label{fig:llm_apt_taxonomy}
\end{figure}

This paper aims to classify LLM-based APT detection approaches in 6G networks and provide a comprehensive taxonomy. Figure \ref{fig:llm_apt_taxonomy} shows this taxonomy and can be summarized in five dimensions:
\begin{itemize}

\item  \textbf{Input Modalities}: LLM models can be fed from various data sources such as logs and PCAP.

\item  \textbf{Detection Granularity}: LLMs can perform APT detection at different levels, such as single-packet analysis and session-based modeling.

\item  \textbf{LLM Techniques}: LLM models can be trained in various ways (prompt tuning, etc.) according to different scenarios.

\item  \textbf{Deployment Models}: LLMs can be deployed on different platform environments, such as cloud computing and edge computing.

\item  \textbf{Threat Lifecycle Phase}: LLM models can provide analysis and interventions at various stages of the APT kill chain.

\end{itemize}

\section{REVIEW METHODOLOGY} \label{sec:Methodology} 

This section presents the methodologies employed in the systematic review of LLM-driven APT detection approaches within 6G wireless networks and subsequently outlines the formulated research questions.

\subsection{Papers Collection}

Since LLM-focused APT detection approaches in 6G wireless networks are a very current topic, we targeted the years 2018-2025 (current) to collect the relevant current literature studies. The following keywords were used to identify the studies related to the research topic:

\begin{enumerate}
  \item \verb|[(LLM) || (LargeLanguageModel)] \& [(APT) || (AdvancedPersistentThreat)]|
  \item \verb|[(6G) || (WirelessNetworks)] \& [(LLM) || (APTDetection)] \& [(Edge) || (CrossLayerSecurity)]|
  \item \verb|[(CyberThreatIntelligence) || (ProvenanceLogs)] \& [(LLM) || (APT)] \& [(6G)]|
  \item \verb|[(LLM)] || [(APT)] || [(6G)]|
\end{enumerate}

\begin{figure}[ht]
	\centering
    \includegraphics[width=0.5\textwidth]{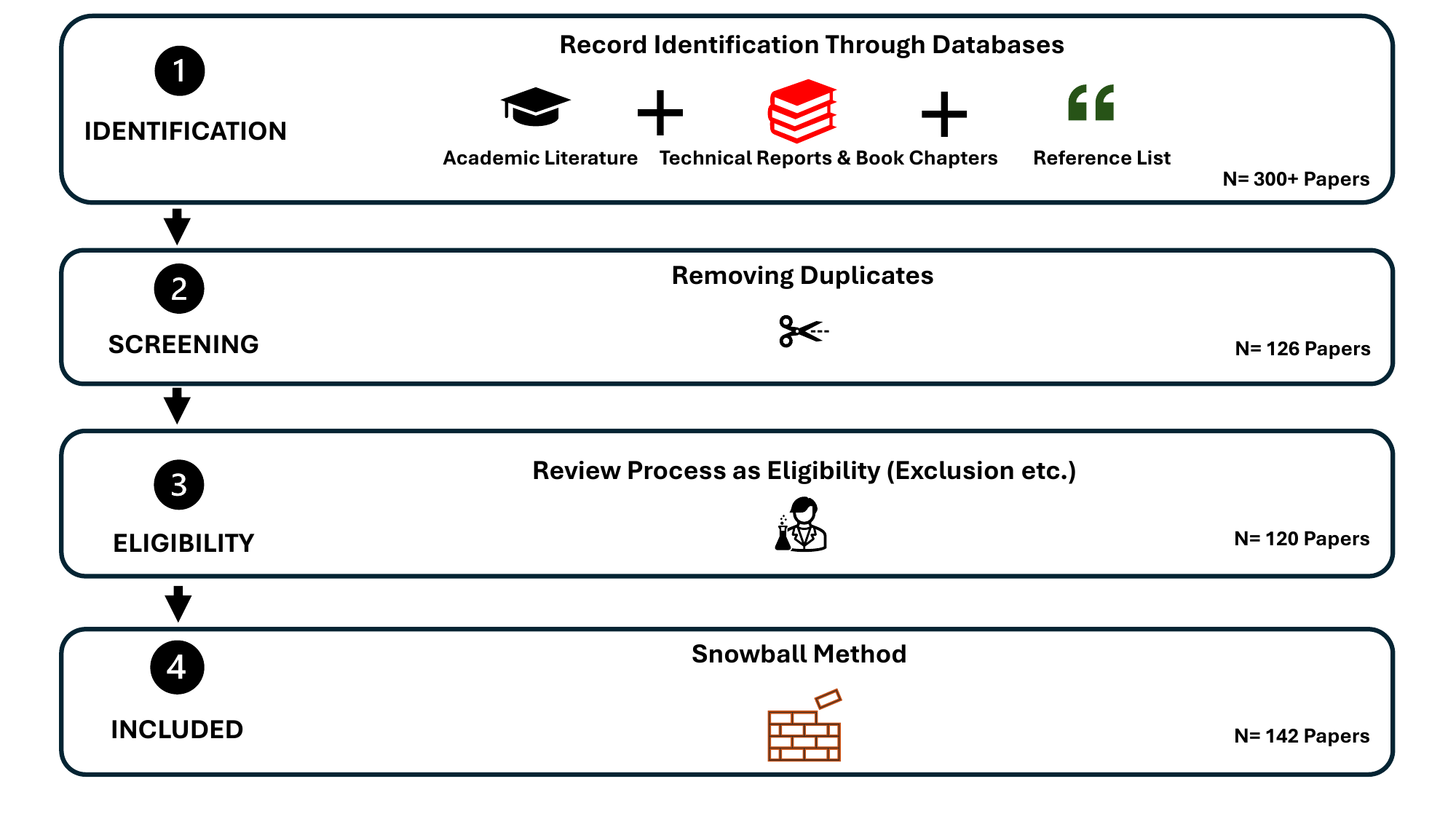}
	\caption{The Paper Collection Process}
	\label{fig:papercollect}
\end{figure}

Figure \ref{fig:papercollect} shows the collection and filtering process of the articles examined in this study. The steps carried out for this process are aimed at providing a comprehensive and structured analysis by following Kitchenham's Systematic Literature Review (SLR) and Petersen's Systematic Mapping Study (SMS) approaches \cite{kitchenham2009systematic,petersen2008systematic}. The steps summarizing this process are explained below:

\begin{itemize} 

\item  \textbf{Identification}: Known major academic literature sources (IEEE, ACM, Elsevier, Springer), technical reports, book chapters, and reference lists were scanned.

\item  \textbf{Screening}: Duplicate documents were removed, and the number of papers decreased to 126.

\item  \textbf{Eligibility}: Papers collected by our expert authors were analyzed, and only quality and scope-compliant papers were selected (the number of papers decreased to 120).

\item  \textbf{Included}: Additional relevant studies were added using the backward and forward snowball method, and the paper set was determined as 142 \cite{golec2024cold}.

\end{itemize}

\subsection{Research Questions} \label{sec:RQ}

The research questions used in the systematic review and the section in which they are examined are shown in Table \ref{tab:research_questions}. Using these research questions, the current literature is examined and analyzed.






\begin{table*}[ht]
\centering
\caption{Summary of Research Questions (RQs), Motivations, and Corresponding Sections}
\label{tab:research_questions}
\resizebox{\textwidth}{!}{%
\begin{tabular}{@{}cccc@{}}
\toprule
\textbf{NO} & \textbf{RQ}  & \textbf{Motivation}& \textbf{Section} \\ \midrule
1           & \begin{tabular}[c]{@{}c@{}}How can LLMs semantically use \\ fragmented provenance data from 6G sources in APT detection?\end{tabular}                                              & \begin{tabular}[c]{@{}c@{}}The aim of this RQ is to investigate how \\ fragmented resource records can be attributed by LLMs in 6G.\end{tabular}                                                   & 5.1              \\
2           & \begin{tabular}[c]{@{}c@{}}What are the limitations of encrypted 6G channels \\ in APT detection and what features of LLM can solve this problem?\end{tabular}                     & \begin{tabular}[c]{@{}c@{}}The aim of this RQ is to examine the \\ potential challenges posed by 6G networks in APT \\ detection and how LLMs can solve them.\end{tabular}                         & 5.2              \\
3           & \begin{tabular}[c]{@{}c@{}}What are the resource constraints in LLMs deployment \\ to 6G edge devices and what techniques can be used to mitigate these limitations?\end{tabular} & \begin{tabular}[c]{@{}c@{}}The aim of this RQ is to explore which compression strategies \\ can be applied when deploying LLMs models to 6G edge devices.\end{tabular}                             & 5.3              \\
4           & \begin{tabular}[c]{@{}c@{}}What are the datasets and modeling methods available for \\ LLM-driven APT detection approaches within 6G wireless networks?\end{tabular}               & \begin{tabular}[c]{@{}c@{}}The aim of this RQ is to investigate suitable datasets and dataset generation \\ methods for LLM-focused APT detection approaches in 6G wireless networks.\end{tabular} & 5.4              \\
5           & \begin{tabular}[c]{@{}c@{}}Where are existing LLM-based APT studies \\ published and do they support reproducibility?\end{tabular}                                                 & \begin{tabular}[c]{@{}c@{}}The aim of this RQ is to evaluate the reproducibility of the \\ dataset and model usability of the reviewed studies.\end{tabular}                                       & 5.5              \\ \bottomrule
\end{tabular}%
}
\end{table*}

\section{ANALYSIS}\label{sec:analysis}

In this section, we discuss how we addressed the research questions in this study.

\subsection{Semantic Correlation of Fragmented Provenance Logs in 6G (RQ1)}

\begin{figure}[ht]
\centering
\includegraphics[width=0.5\linewidth]{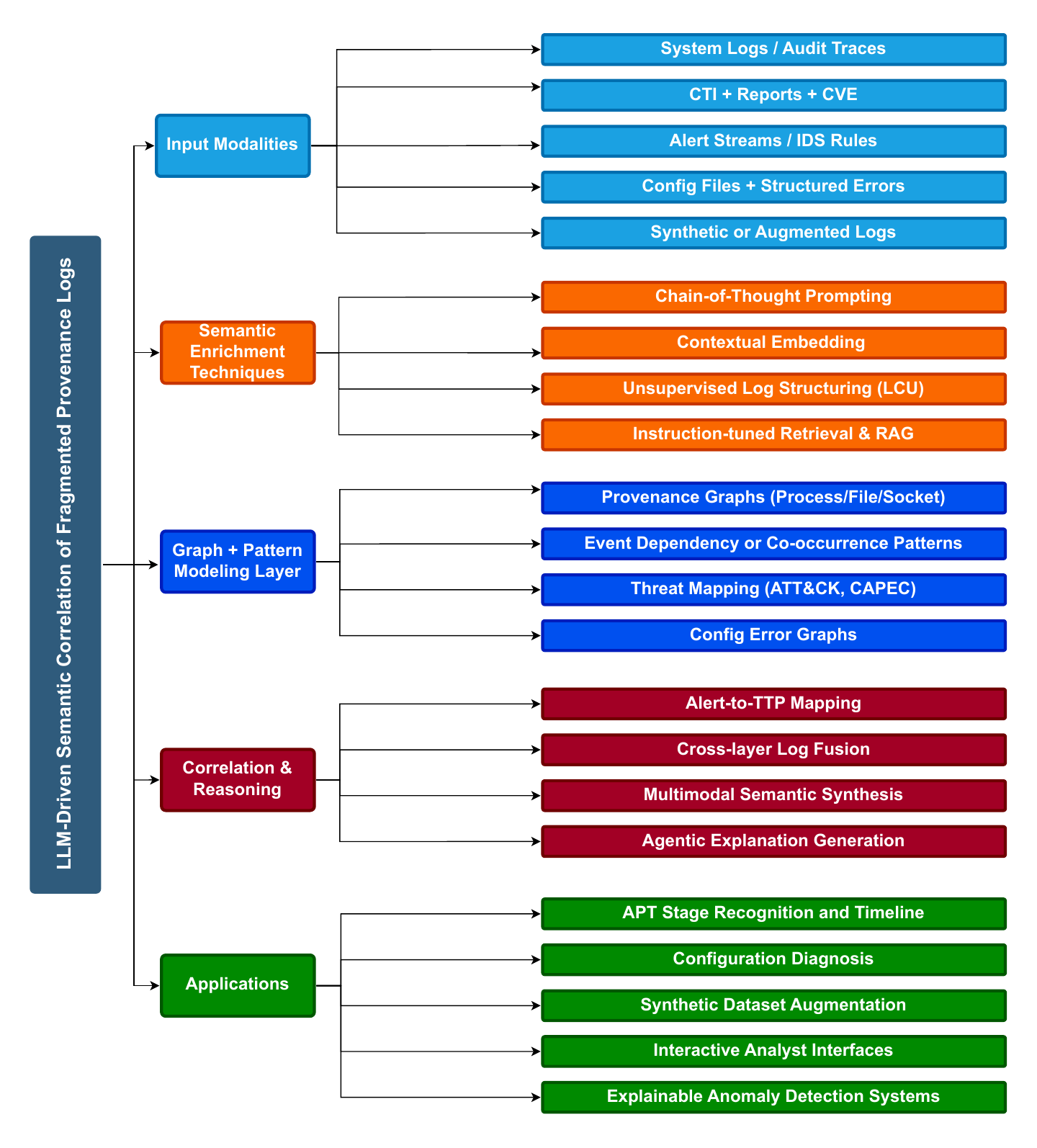}
\caption{RQ1 Taxonomy: LLM-based Semantic Correlation of Fragmented Provenance Data Across Heterogeneous 6G Sources}
\label{fig:rq1-taxonomy}
\end{figure}

6G networks contain a lot of fragmented lineage data due to their heterogeneous structure (edge, cloud, etc.), and this data is distributed and inconsistent, which causes difficulties for security analysts and attack systems in APT detection \cite{sergiou2020complex}. One example of these difficulties is that rule-based and statistical detection methods fail to capture the nuanced context required for attack detection \cite{ sood2025hallucination, alturkistani2025slr}. Recent studies have focused on LLM-based methods to semantically combine fragmented lineage data and provide context-aware correlation. Figure \ref{fig:rq1-taxonomy} shows how fragmented lineage data can be semantically associated with LLM-enabled systems in a multi-layered manner. LLMs offer a promising solution by generating consistent security narratives by syntactically and temporally handling various records (such as security logs) \cite{benabderrahmane2025apt, gandhi2025shield, he2024llmelog, shan2024face, arikkat2025droidttp}.

Recent findings have shown that LLM's models can effectively utilize many different sources, such as audit logs \cite{benabderrahmane2025apt}, IDS alerts \cite{g2024harnessing}, CTI reports \cite{zhang2024tactics}, and even static code artifacts \cite{ignatyev2024large}. Models that transform low-level source sequences into textual formats, such as APT-LLM \cite{benabderrahmane2025apt}, GENTTP \cite{zhang2024tactics}, and LLMeLog \cite{he2024llmelog}, have been developed to reflect the system behavior semantics of models such as BERT or RoBERTa. Frameworks based on multitasking instructions and thought chains, such as SEVENLLM \cite{ji2024sevenllm} and AnomalyGen \cite{li2025anomalygen}, have been proposed for reasoning in data-scarce environments. For enrichment techniques, the literature includes studies such as retrieval-augmented generation \cite{gandhi2025shield}, clustering embedding \cite{huang2024lunar, he2024llmelog}, and ATT\&CK alignment via request templates \cite{koendersadvancing}.

Many frameworks have been proposed that support fragmented logs with deep reasoning by capturing temporal, causal, and entity-level relationships and that resort to graph-based modeling. SHIELD \cite{gandhi2025shield}, MultiKG \cite{wang2024multikg}, and MAD-LLM \cite{du2024mad} frameworks use source graphs that encode dependencies of edges and represent system events at nodes. Other works such as AURORA \cite{wang2024sands} and DroidTTP \cite{arikkat2025droidttp} reconstruct attack sequences by applying classical planning and LLM. Works such as LocalIntel \cite{mitra2024localintel} and MCM-LLAMA \cite{diakhame2024mcm} prefer dynamic association of SOC information and external alerts, while works such as LUNAR \cite{huang2024lunar} and AnomalyGen \cite{li2025anomalygen} prefer association with CTI corpora. For high-level reasoning and explanation generation, these works resort to semantically annotated graph-based modeling.

Despite all these developments, there are still limitations that remain to be addressed. SEVENLLM \cite{ji2024sevenllm} and SHIELD \cite{gandhi2025shield} frameworks use organized and synthetic logs, but this does not fully reflect the dynamic, heterogeneous nature of 6G. Another point to note is that mitigation strategies such as hybrid verification \cite{benabderrahmane2025apt} and instruction fine-tuning \cite{ji2024sevenllm} are rarely applicable to edge contexts. In addition, LLMs' high processing power and storage requirements make their application in 6G edge nodes a serious challenge, and therefore, the need for lightweight alternative methods such as TinyLM agents \cite{yang2024tinyhelen} or MoE-based distributed inference \cite{du2024mad} is increasing. For future research, areas such as cross-layer lineage fusion, real-time semantic timeline reconstruction, and hallucination-aware causal modeling \cite{ ignatyev2024large} stand out as a research gap.

\subsection{Limitations of Encrypted 6G Channels and LLM-Driven Solutions (RQ2)}

\begin{figure}[ht]
\centering
\includegraphics[width=0.5\linewidth]{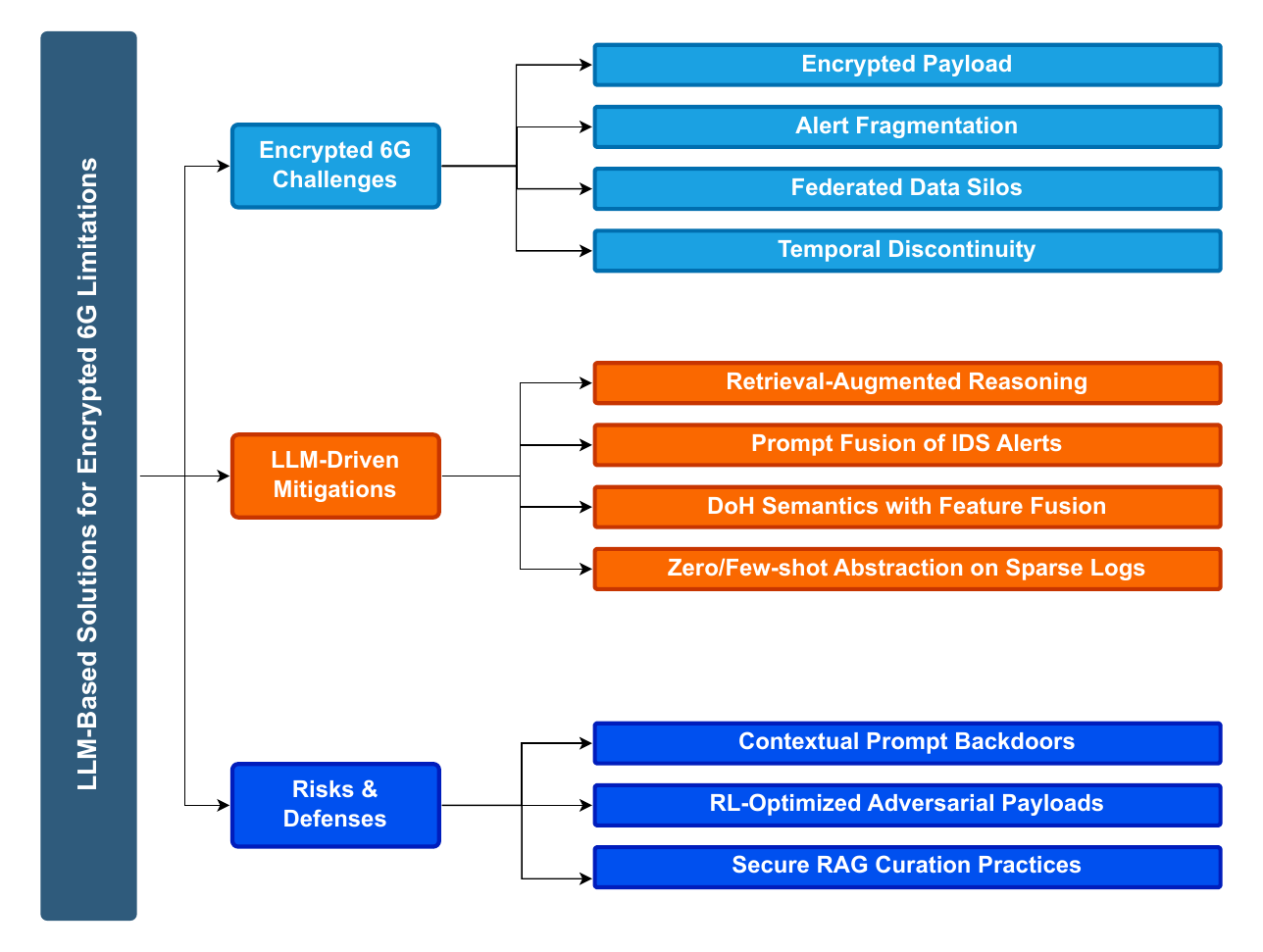}
\caption{RQ2 Taxonomy: LLM-Based Solutions for Encrypted 6G Limitations}
\label{fig:llm_tree_encryption}
\end{figure}

The widespread use of some communication protocols, such as DNS-over-HTTPS (DoH) and end-to-end encrypted tunnels, in the transition to 6G wireless networks has made great contributions to security and user privacy. In addition to these contributions, it also brings disadvantages like blind spots, such as traffic semantics obscurity for AI-supported detection systems. Figure~\ref{fig:llm_tree_encryption} shows a taxonomy of LLM-focused solutions offered to address the challenges, limitations, and risks of 6G networks due to encrypted channels.

\subsubsection{Technical Limitations Imposed by Encryption}

Encrypted 6G traffic channels limit the visibility of attack surfaces due to the techniques used (such as DoH). Recent studies have shown that advanced DL models fail to detect malicious traffic because semantic payloads become ambiguous while being encrypted \cite{sun2025advtg}. In addition, advanced attack methods such as APT try to avoid detection by using encrypted channels such as DoH and embedding the C2 infrastructure in HTTPS payloads \cite{diao2024poster}. Edge-based data isolation, whose main purpose is privacy, prevents correlation (temporal and spatial) between devices. For example, since fragmented traffic logs are produced in UAV-based 6G networks, anomaly monitoring becomes very difficult \cite{hadi2024real}

\subsubsection{LLM-Driven Mechanisms to Address These Gaps}

To overcome all these limitations, LLMs are promising by making meaningful inferences with their capabilities in semantic reasoning and contextual abstraction.

A recent study, APTSniffer, is a framework that detects APTs in encrypted channels by converting flow features into textual prompts \cite{xu2025aptsniffer}. The results confirm that the framework is successful with a 97\% F1 score. Another study, MAD-LLM, is a framework that reconstructs APT chains by semantically collecting them through LLMs despite fragmented IDS alerts and encryption at the network layer \cite{du2024mad}

Some malware (such as DoHunter, Godlua) are difficult to detect by detection systems because they use encrypted channels, so researchers track some technical features of the traffic, such as timing, length, and target domain structure, in addition to raw data with LLM models \cite{diao2024poster}.

\subsubsection{Emerging Challenges and Threats}

Although using LLM models in encrypted channels is a promising solution, it is important to consider LLM-based vulnerabilities. One of these vulnerabilities is that LLM behavior can be manipulated by hostile requests and poisoned rollbacks. Studies have confirmed that fine-tuned LLM models based on RL can generate malicious traffic \cite{sun2025advtg}. Furthermore, literature confirms that LLM models inject hidden logic into LLM models that are activated by benign triggers in encrypted channels \cite{liu2025compromising}

In conclusion, while LLM models provide an advantage, such as semantic visibility for attack detection in encrypted channels, they also inherently introduce attack surfaces.

Table~\ref{tab:llm1} summarizes the main limitations of encrypted 6G environments in light of the current literature reviewed.

\begin{table}[ht]
\centering
\caption{Mapping Encrypted 6G Challenges to LLM-Driven Solutions}
\label{tab:llm1}
\resizebox{\linewidth}{!}{%
\begin{tabular}{@{}lll@{}}
\toprule
\textbf{Work}        & \textbf{Limitation}                            & \textbf{LLM-Based Technique}                                      \\ \midrule
Xu et al. \cite{xu2025aptsniffer}         & Payload Obfuscation                            & Retrieval-Augmented Inference                                     \\
Du et al. \cite{du2024mad}            & Alert Fragmentation                            & Multi-stage Reasoning via Prompt Engineering                      \\
Diao et al. \cite{diao2024poster}          & Covert DoH C2 Channels                         & LLM + Expert Features for Tunnel Detection                        \\
Cheng et al. \cite{cheng2025sok}         & Contextual Reasoning in Sparse Logs            & Log Fusion and Interpretation via Few-shot Learning               \\
Sun et al. \cite{sun2025advtg}          & LLM Model Poisoning via Traffic                & Adversarial Sample Generation with Reinforcement Learning (RL)    \\
Liu et al. \cite{liu2025compromising}          & Contextual Logic Corruption                    & In-context Backdoor Prompt Manipulation                           \\ \bottomrule
\end{tabular}%
}
\end{table}

\subsection{Deploying LLMs at the Edge: Constraints and Optimization Techniques (RQ3)}

\begin{figure}[ht]
\centering
\includegraphics[width=0.5\linewidth]{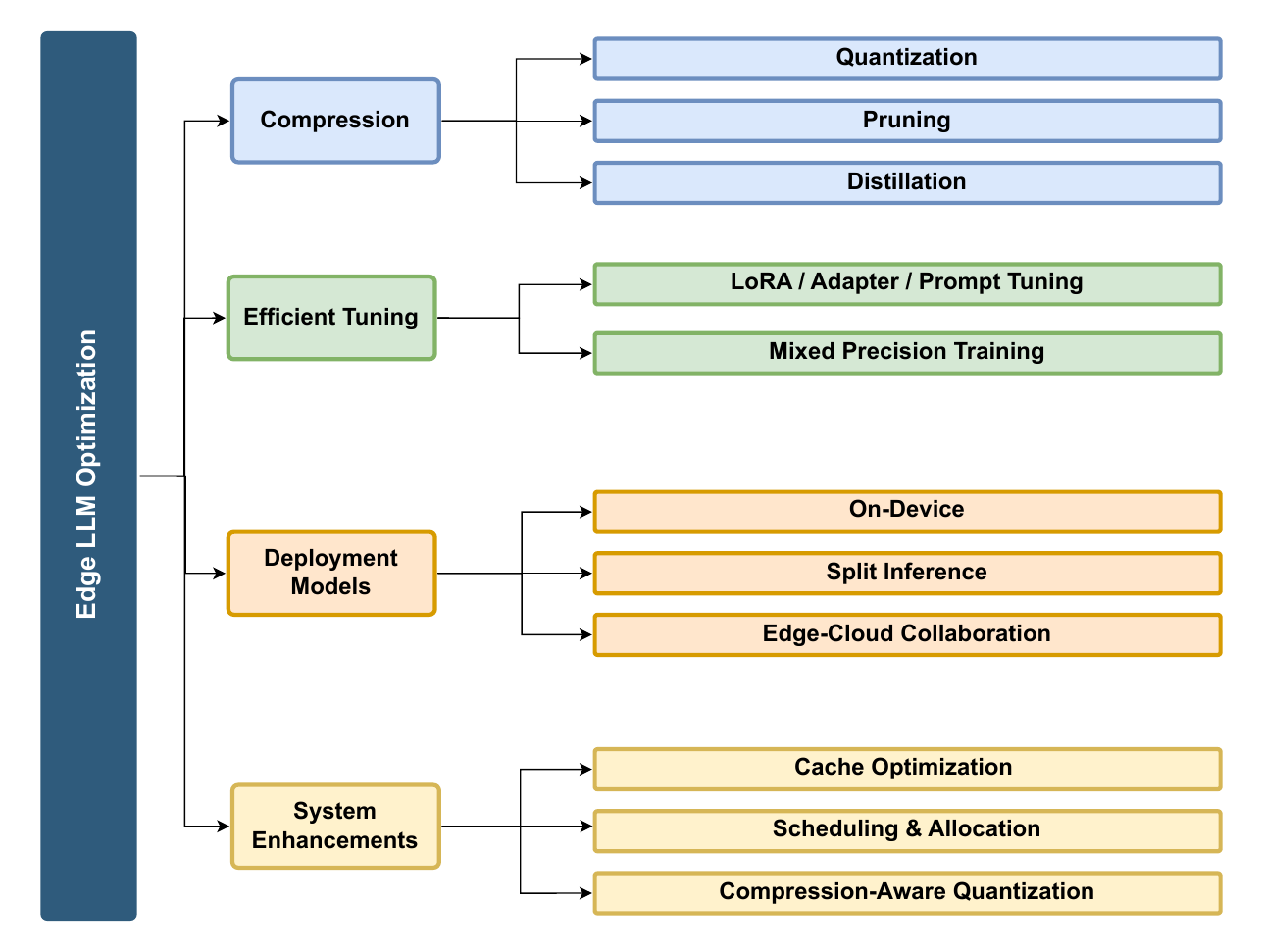}
\caption{RQ3 Taxonomy: Edge-Oriented LLM Optimization Strategies}
\label{fig:edge_llm_taxonomy}
\end{figure}

Despite the advantages of high speed and low latency offered by 6G networks, they consist of many different distributed nodes and heterogeneous structures, such as edge devices. Therefore, LLM models to be used for security, privacy, and context-adaptive smart applications should also take into account the major computational, architectural, and security-related challenges when deployed in 6G networks. This research question (RQ3) examines optimization techniques for edge scenarios by categorizing these constraints. Figure \ref{fig:edge_llm_taxonomy} shows edge-oriented LLM optimization strategies.

\textbf{Resource Constraints in Edge Environments}: Edge devices (IoT, smartphones, etc.) consist of devices with limited processing power and storage capabilities, and even very small LLM models require more than 7GB RAM, which is usually beyond the capacity of edge computing devices \cite{wang2025compression}. Furthermore, due to the nature of LLMs (autoregressive), sequential token generation may cause latency bottlenecks \cite{zhao2024edge}.

\textbf{Security and Fairness Considerations}: Since LLM models deployed on edge nodes typically handle user data, privacy concerns may arise if this data is compromised \cite{khowaja2024pathway}. Additionally, recent studies have reported that compression techniques used to make LLM models lighter for edge nodes may increase bias against underrepresented groups \cite{xu2024beyond}. Therefore, the issue of fairness and reliability in compression cases is an open research gap.

\textbf{Model Compression Techniques}: These are techniques used to reduce the memory and processing load of LLM models, and one of the most popular methods is quantization. In this method, the weights (such as FP32) are converted to lower bit representations (INT8 or FP4) to reduce the model size and optimize the hardware speed \cite{wang2024model,liu2024designing}. Another method where the distribution is optimized is pruning, and in this method, the weights are rescaled before quantization \cite{wang2025compression}. Distillation and low-rank approximation methods aim to provide additional performance gains on the inference quality \cite{xu2024beyond,wang2024model}.

\textbf{Parameter-Efficient Fine-Tuning (PEFT)}: Fine-tuning of LLM models cannot be done at edge nodes, and therefore PEFT methods (such as LoRA, Adapters, and Prompt-Tuning) apply them to local tasks by updating a subset of the models' parameters \cite{qin2024empirical}. Another recent research introduces a new collaborative training model that optimizes fine-tuning of early layers at the edge device (mobile) and deep layers at the edge server \cite{liu2024resource}. In this model, the aim is to reduce communication and energy costs while keeping the personalized performance constant.

\textbf{Collaborative and Split Inference}: Another rational approach is to distribute the LLM model across the device-edge-cloud heterogeneity to achieve the balance of performance and resource utilization. Yang et al. \cite{yang2024perllm} propose a structure in which the cloud and edge are jointly used to offload LLM inference with a UCB-based scheduler. The results show that the energy usage is halved and the efficiency is doubled. Another approach is to share the converter layers among the heterogeneous edge devices using the matching theory \cite{picano2025matching}

\textbf{System-Level Runtime Optimizations}: Another method of increasing efficiency is runtime and architecture-based approaches:

\begin{itemize}

\item  \textbf{KV-cache compression}: It is the process of reorganizing memory to save RAM on edge devices \cite{liu2024designing}

\item  \textbf{Contextual sparsity and batch-aware scheduling}: In contextual sparsity, the process of reducing the processing load by looking only at the important tokens of the model, while in batch-aware scheduling, the process of running multiple tasks in the most efficient order without blocking each other \cite{liu2024resource, liu2024designing}.

\item  \textbf{Speculative decoding}: It is the process of predicting multiple tokens at the same time, and the aim is to reduce the autoregressive latency bottleneck \cite{zhao2024edge}.

\end{itemize}

Table \ref{tab:llm_edge_optimization} summarizes the main optimization techniques developed against resource constraints on edge devices, their usage scenarios, and the cost/benefit balances they bring.

\begin{table*}[ht]
\centering
\caption{Optimization Techniques for Deploying LLMs under Edge Constraints}
\label{tab:llm_edge_optimization}
\resizebox{\textwidth}{!}{%
\begin{tabular}{@{}ccccc@{}}
\toprule
\textbf{Reference}     & \textbf{Technique}             & \textbf{Constraint Addressed} & \textbf{Use Case}            & \textbf{Trade-off}       \\ \midrule
Wang et al. \cite{wang2024model}  & INT4/INT8 Quantization         & Memory, Compute               & Mobile Edge Inference        & ↓Accuracy, ↑Speed        \\
Wang et al. \cite{wang2025compression}  & Compression-aware Quantization & Memory, Latency               & Smartphone + AI Assistant    & Low overhead             \\
Qin et al. \cite{qin2024empirical}   & LoRA / Adapters                & Fine-Tuning cost, Storage     & Personalized edge assistants & ↓Flexibility, ↑Privacy   \\
Yang et al. \cite{yang2024perllm}  & PerLLM Scheduler               & QoS-aware Scheduling          & Edge-Cloud Mixed Load        & ↑Efficiency, ↑Throughput \\
Zhao et al. \cite{zhao2024edge}  & Token Parallel Decoding        & Latency (token gen)           & Edge-terminal co-inference   & Complex sync             \\
Picano et al. \cite{picano2025matching} & Matching-based Layer Placement & Device Heterogeneity          & Heterogeneous Edge Inference & ↑Accuracy                \\ \bottomrule
\end{tabular}%
}
\end{table*}

\subsubsection{Federated or Distributed Training of LLMs at the Edge}

With the proliferation of user-centric applications, it is expected that deployment strategies of LLM models will be developed on edge devices as well \cite{liu2025compromising}. Cloud-based systems cause additional latency and bandwidth loads in 6G environments compared to edge computing \cite{ferrag2023edge}. Therefore, Federated Learning (FL) and distributed tuning paradigms can be used to reduce these loads by processing data on edge devices \cite{ferrag2023edge}.

\textbf{Motivation for Federated Edge Training}: 

Compared to traditional cloud-based systems, training on edge devices provides improvements in the following limitations \cite{khowaja2024pathway, qu2025mobile, zhao2024edge}:

\begin{itemize}
  
\item  \textbf{Privacy}: Since sensitive data, such as biometric data, is processed on edge devices, there are fewer privacy concerns than cloud systems that use central servers.

\item  \textbf{Latency}: Since data is processed close to the data source, there is less latency than cloud-based systems.

\item  \textbf{Bandwidth}: Since cloud-based systems are used only for operations that require large processing power, unnecessary communication bandwidth is not used.
 
\end{itemize}

\textbf{Federated Fine-Tuning Techniques for LLMs}:

In constrained environments (such as communication and computation), LLM model personalization schemes can be summarized as follows:

\begin{itemize}
    
\item  \textbf{Parameter efficient}: In LoRA-based work, low-rank matrices are fine-tuned among clients to reduce transmission volume \cite{liu2024resource}.

\item  \textbf{Split Federation Learning}: Qu et al. \cite{qu2025mobile} propose to train the first layers of the model on the device and optimize the deep layers on edge nodes in their proposed framework called Mobile Edge Intelligence (MEI).

\item  \textbf{Inter-device gradient fusion}: In order to dynamically balance the update frequency and energy budgets, distributed scheduling algorithms are proposed in \cite{zhao2024edge}.

\end{itemize}

\textbf{Challenges in Federated LLM Training}: 

 Despite the advantages of low latency and low bandwidth overhead, federated training at the edge also suffers from statistical heterogeneity \cite{wang2024model}, system heterogeneity \cite{zhang2024edgeshard}, and security risks \cite{khowaja2024pathway}. To overcome these challenges, recent research focuses on model-system co-design based techniques. These techniques include adaptive aggregation \cite{yang2024perllm} where clients are weighted according to their trust scores, compression-aware updates \cite{wang2025compression} where updates are sparse before transmission, and energy-aware scheduling \cite{liu2024designing} where the training frequency is dynamically adjusted to preserve battery and network life. Table \ref{tab:llm2} provides a comparative overview of federated and distributed education strategies.

\begin{table}[ht]
\centering
\caption{Federated and Distributed LLM Training: Constraints and LLM-Based Trade-offs}
\label{tab:llm2}
\resizebox{\linewidth}{!}{%
\begin{tabular}{@{}lll@{}}
\toprule
\textbf{Work} & \textbf{Constraint Addressed} & \textbf{LLM-Based Technique / Trade-off} \\ \midrule
Liu et al. \cite{liu2024resource}     & Bandwidth, Memory        & LoRA-Based FL for Lightweight Personalization (Lower Global Accuracy) \\
Qu et al. \cite{qu2025mobile}        & Compute Offloading       & Split Learning (MEI4LLM) with Multi-layer Collaboration (Sync Overhead) \\
Zhao et al. \cite{zhao2024edge}      & Latency, Energy          & Parallel Token Learning at Edge-Terminals (Complex Scheduling) \\
Qin et al. \cite{qin2024empirical}   & Fine-tuning Privacy      & Federated Prompting + PEFT (Bias Sensitivity in User-Tuning) \\
Yang et al. \cite{yang2024perllm}    & Device Reliability        & Trust-Aware Aggregation in FL (Risk of Model Divergence) \\ \bottomrule
\end{tabular}%
}
\end{table}

\subsection{Datasets and Modeling Techniques for LLM-Driven APT Detection (RQ4)}

\begin{figure}[ht]
  \centering
  \includegraphics[width=0.5\textwidth]{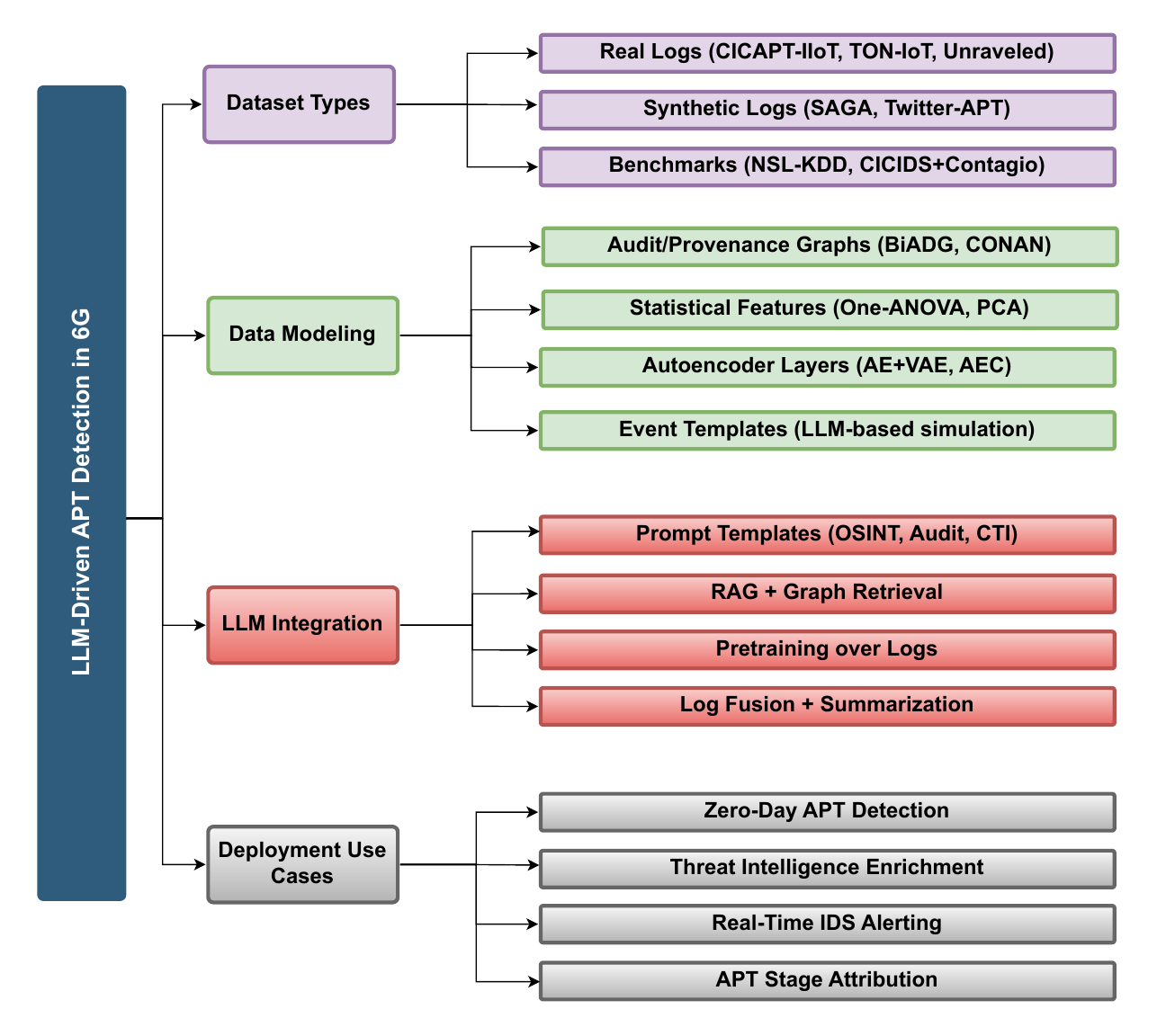}
  \caption{Taxonomy of Dataset–Model–LLM Alignment in 6G APT Detection Pipelines}
  \label{fig:taxonomy_tree}
\end{figure}

The quality of the datasets to be used to train models in LLM-based APT detection in 6G networks directly affects the success rate. The datasets created as a result of examining 32 different studies and the results of the systematic and taxonomy study on modeling techniques are examined in this subsection. Figure \ref{fig:taxonomy_tree} shows this taxonomy and its subsections.

\textbf{Dataset Types for LLM-Based APT Detection}: When the literature is examined, it is seen that APT datasets can be examined under three main headings:

\begin{itemize}

\item \textbf{Semi-Synthetic Datasets}: Semi-synthetic datasets that model APT attacks are as follows: (i) Unraveled dataset \cite{myneni2023unraveled}, which combines real cloud infrastructure logs and simulated APT stages, and (ii) edge-based CICAPT-IIoT dataset \cite{ghiasvand2024cicapt}, which includes UAV and ICS smart environment logs.

\item \textbf{Synthetic and Augmented Logs}: Synthetic datasets that model APT attacks are as follows: (i) SAGA \cite{huang2024saga}, which consists of audit logs compatible with the ATT\&CK matrix, (ii) Twitter-APT \cite{shafee2024evaluation}, which is created by applying LLM's to OSINT-based threats, (iii) a dataset where labeled attacks are created using pcap filters, log records and IDS simulation \cite{al2023machine}.

\item \textbf{Merged Benchmark Corpora}: For APT detection, datasets that are based on a real organization's network traffic and model attacks such as trojans and spyware can be used \cite{neuschmied2022apt}.

\end{itemize}

NSL-KDD or CICIDS data outside these categories are now outdated and fail to model real APTs (stealth lateral movement or long-term dormancy strategies) \cite{oleiwi2023meta, saeed2023anomaly}

\textbf{Data Modeling Techniques and Representations}: Data modeling strategies that can be used to combine 6G network data with LLM models can be summarized as follows:

\begin{itemize}

\item \textbf{Behavioral Graph Profiling}: In this modeling method, BiADG and MIG models are obtained by applying Graph Convolutional Network (GCN) on IP flow graphs and behavior patterns \cite{xuan2024novel, nguyen2023new}. In addition, there is the CONAN model that provides low-latency matching for APT stages using a Finite State Machine (FSM) \cite{xiong2020conan}.

\item \textbf{Statistical + Feature Engineering Pipelines}: As an example of these strategies, two separate studies that apply preprocessing such as One-ANOVA based cleaning, decomposition, and boosting by synthetic generation \cite{neuschmied2022apt, al2023machine} can be given as examples.

\item \textbf{Multi-Stage Autoencoders}: In APTSID \cite{neuschmied2022apt}, where this strategy is applied, standard and variational autoencoders are combined with statistical feature selection to achieve high accuracy anomaly detection.

\item \textbf{ML + Expert System Hybrids}: In the CDT system \cite{al2023machine}, where this technique is used, an attack detection prediction is taken with an ML model and transmitted to the rules used by systems such as SMORT.

\end{itemize}

Table \ref{tab:dataset_model_comparison} shows the comparison of datasets and modeling techniques in LLM-Driven APT detection studies

\textbf{LLM Integration Strategies}: These are the methods used when integrating LLM models into various systems, and the main purpose is to enable LLM models to be used with various data.

\begin{itemize}

\item \textbf{Prompt Templates + Simulation}: LLM prompts are the methods used to generate attack data and multiply training data, and SAGA and CyExec are two examples of this in academia \cite{huang2024saga, yamin2024applications}.

\item \textbf{OSINT + NER Pipelines}: Although LLM models are successful in detecting threats in open source articles, fine-tuning is required for small details. Shafee et al.\cite{shafee2024evaluation} tries to find threats from open source information with LLM.

\item \textbf{Fusion Architectures}: In these methods, after the data is processed with other models and made meaningful, it is given to the LLM model to process, and thus it is expected that LLM will perform a more successful analysis. Models such as AE+VAE and AE-CNN use this method \cite{neuschmied2022apt}

\end{itemize}

\begin{table*}[ht]
\centering
\caption{Comparison of Datasets and Modeling Techniques in LLM-Driven APT Detection Studies}
\label{tab:dataset_model_comparison}
\resizebox{\textwidth}{!}{%
\begin{tabular}{@{}ccccc@{}}
\toprule
\textbf{Reference}         & \textbf{Dataset}    & \textbf{Modeling Technique}      & \textbf{LLM Use}          & \textbf{Key Insight}                                                                               \\ \midrule
Huang et al. \cite{huang2024saga}      & SAGA (Synthetic)    & Prompt-Based Log Generation      & Training Input            & \begin{tabular}[c]{@{}c@{}}ATT\&CK-aligned, \\ synthetic audit logs for \\ APT stages\end{tabular} \\
Neuschmied et al. \cite{neuschmied2022apt} & CICIDS + Contagio   & AE + VAE Stack                   & Feature Compression       & \begin{tabular}[c]{@{}c@{}}Multi-stage anomaly detection \\ with zero-day support\end{tabular}     \\
Al-Aamri et al. \cite{al2023machine}   & Custom Logs         & CDT + SNORT Rule Feed            & Manual LLM Friendly     & \begin{tabular}[c]{@{}c@{}}Time-series + journaling logs \\ with rule generation\end{tabular}      \\
Ghiasvand et al. \cite{ghiasvand2024cicapt}  & CICAPT-IIoT         & Provenance + Network Flow Fusion & LLM-Graph Possible        & \begin{tabular}[c]{@{}c@{}}Audit trails + flow logs for \\ IIoT APT detection\end{tabular}         \\
Shafee et al. \cite{shafee2024evaluation}    & Twitter Corpus      & OSINT Classification + NER       & NER and Prompt Evaluation & \begin{tabular}[c]{@{}c@{}}LLMs need domain adaptation \\ for threat-level NER\end{tabular}        \\
Xuan et al. \cite{xuan2024novel}      & BiADG               & Behavioral GCN + LSTM            & Graph-to-LLM Potential    & IP-node behavior modeling over graph structures                                                    \\
Oleiwi et al. \cite{oleiwi2023meta}     & NSL / CICIDS / UNSW & Meta-Model Voting Ensemble       & Pre-LLM Classifier Layer  & Traditional ML stack for high-precision filtering                                                  \\ \bottomrule
\end{tabular}%
}
\end{table*}

\subsection{Reproducibility and Publication Trends in LLM-Based APT Studies (RQ5)}

\begin{figure}[ht]
    \centering
    \includegraphics[width=.5\textwidth]{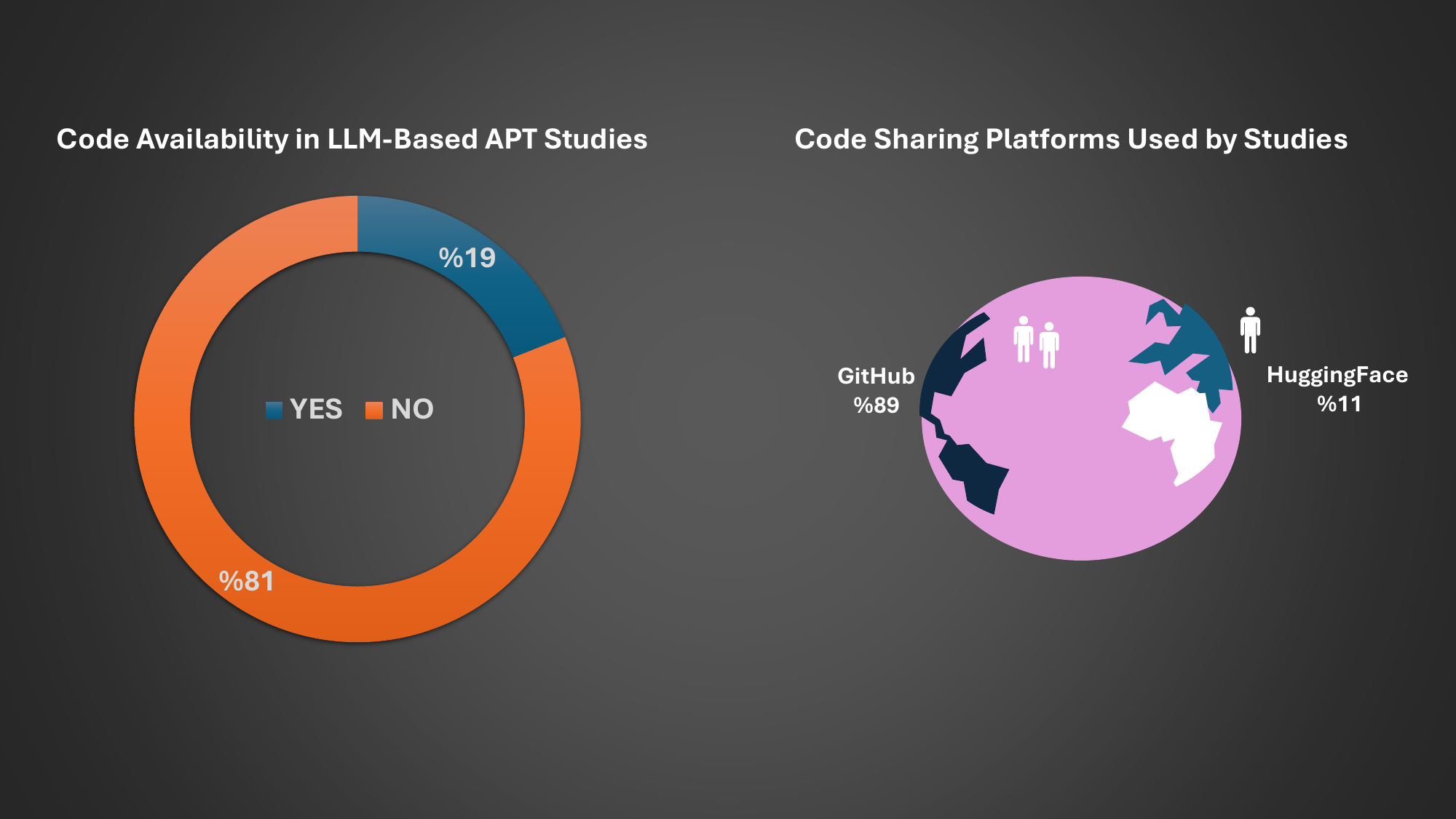}
    \caption{The Percentage of LLM-based APT Studies that Shared Source Code (YES/NO) and the Platforms where the Code was Hosted}
\label{fig:code_availability_platforms}
\end{figure}

This research question questions the reproducibility and other statistical information of LLM-based APT detection studies. In order to provide a comprehensive assessment of the 142 recent studies utilized throughout the paper, we have classified all papers in \textbf{Appendix A} according to \textit{Code Availability}, \textit{Dataset Evaluation}, \textit{Protocol Venue/Platform}, and \textit{Year}. The description of these features and the resulting statistical information are as follows:

\textbf{Code Availability}: This feature was used to classify studies according to their reproducibility. Figure \ref{fig:code_availability_platforms} shows how many percentage of the studies shared their source code (YES/NO), and on which platform (Github, etc.) they were published. As can be seen from the figure, only a very small portion of the examined studies shared their source code, while most of their code was published on the GitHub platform.

\begin{figure}[ht]
    \centering
    \includegraphics[width=.5\textwidth]{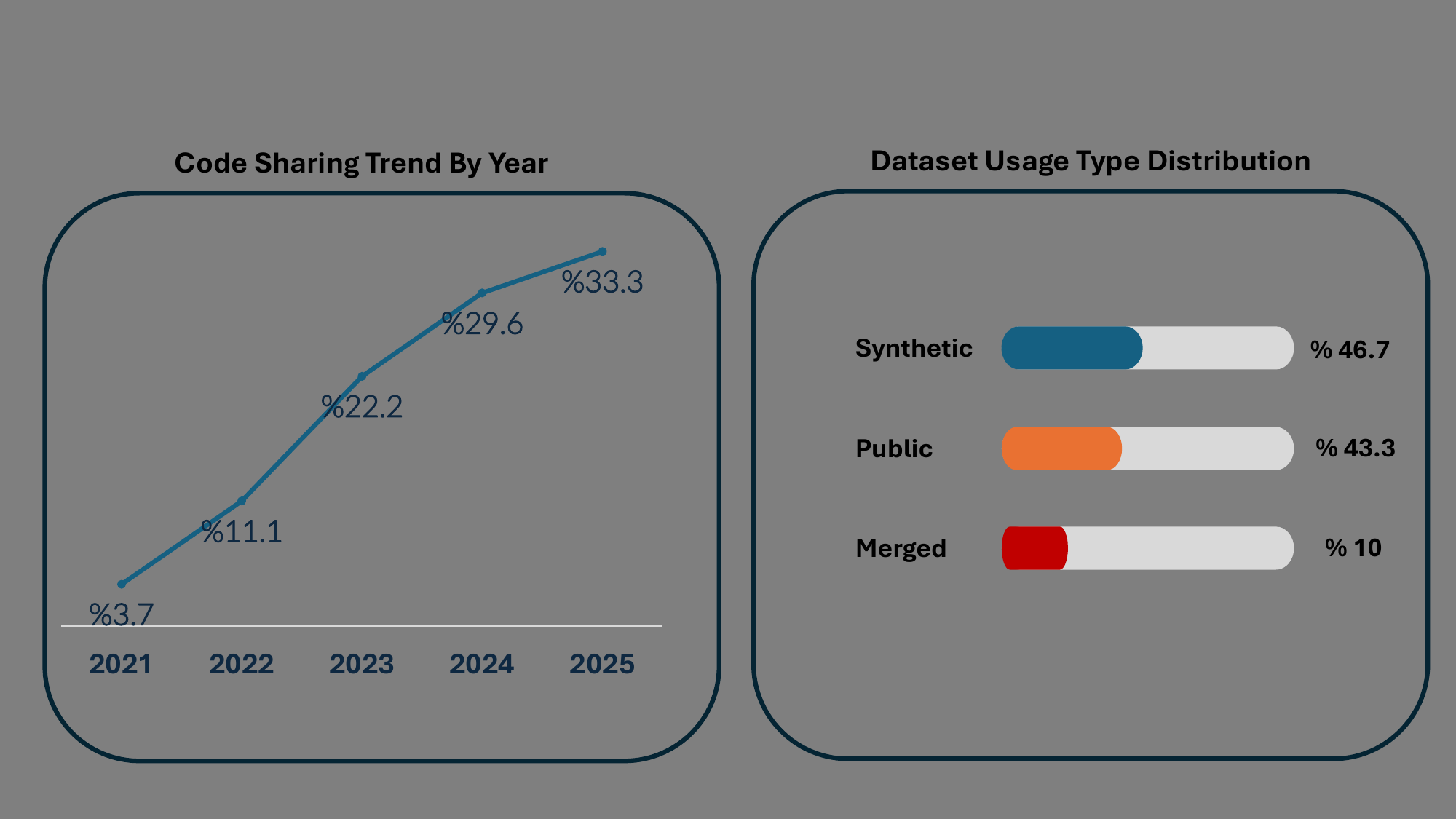}
    \caption{The Trend of Dataset Usage over the Years and Most Commonly used Datasets in LLM-based APT Studies}
    \label{fig:dataset_trends}
\end{figure}

\textbf{Dataset}: This column was used to measure the diversity of datasets used in the studies and to determine how many of them used real-world data. Figure \ref{fig:dataset_trends} shows the percentage of datasets shared by year and the percentage of articles using synthetic-public datasets. The results confirm that datasets used in APT detection studies tend to be shared and that the most used dataset is synthetic dataset.

\begin{figure}[ht]
    \centering
    \includegraphics[width=.5\textwidth]{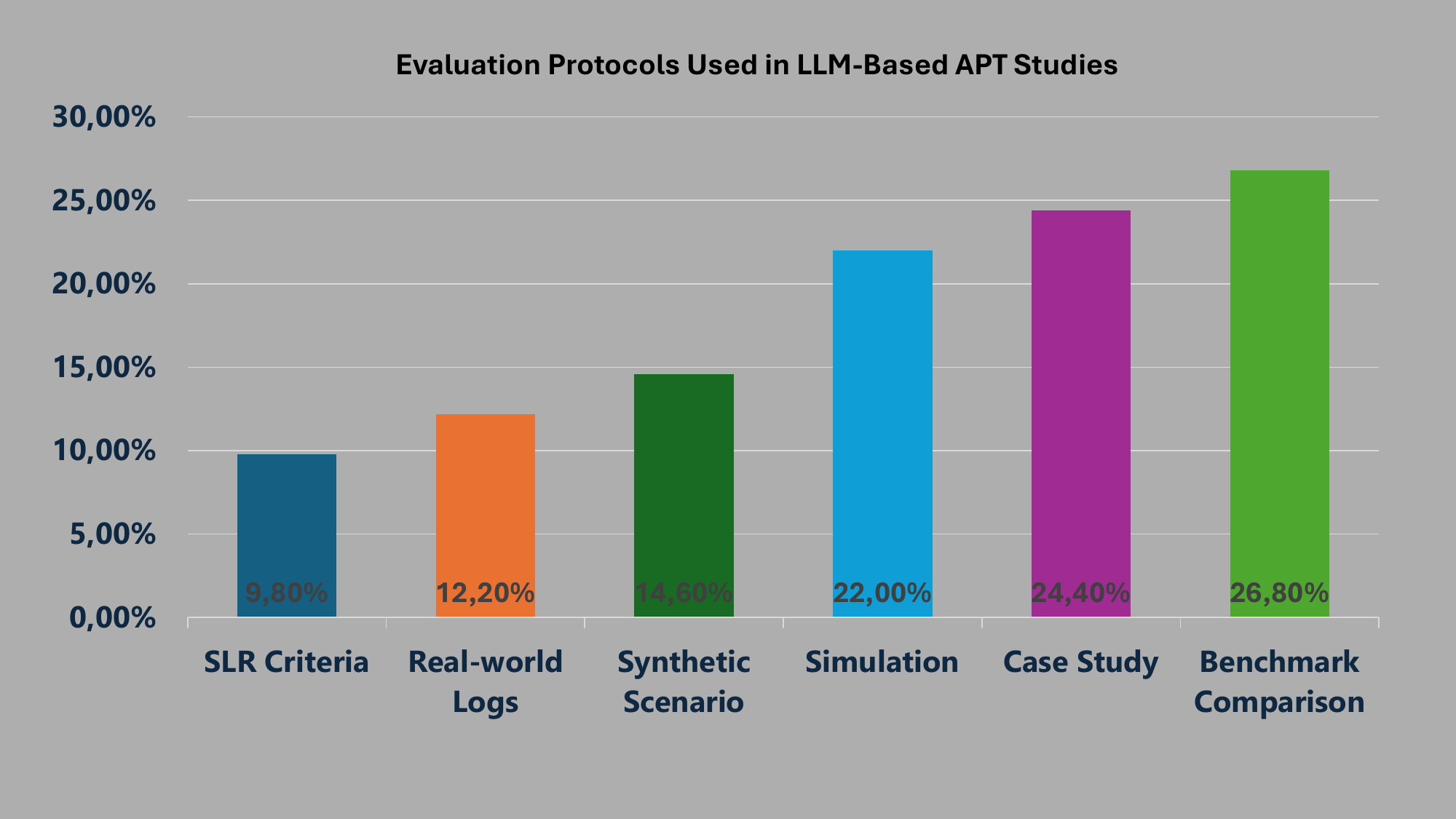}
    \caption{Evaluation Protocols Used in LLM-Based APT Studies}
    \label{fig:evaluation_protocols}
\end{figure}

\textbf{Evaluation Protocol}: This column is to evaluate the level of empirical validity of the reviewed articles based on whether they use robust protocols such as cross-validation. Figure \ref{fig:evaluation_protocols} shows the frequency of the protocols used.

\begin{figure}[ht]
    \centering
    \includegraphics[width=.5\textwidth]{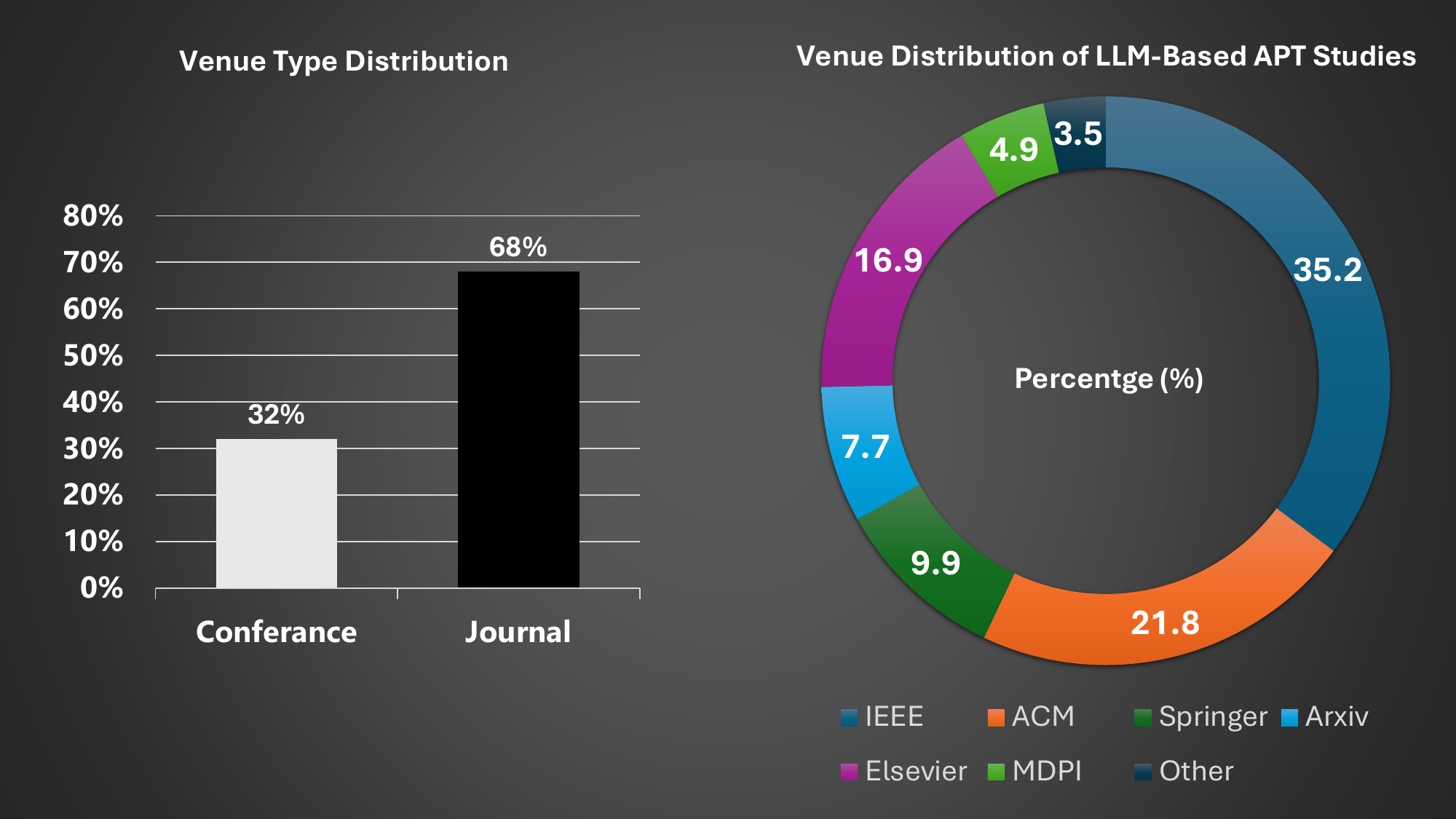}
    \caption{Distribution of Venues where LLM-based APT Studies were Published and Their Types}
    \label{fig:venue_types_distribution}
\end{figure}

\textbf{Venue / Platform}: This column examines the publication quality and field spread by examining the venue/platforms and types (conference/journal) where the reviewed studies were published. Figure \ref{fig:venue_types_distribution} shows a summary of this statistic.

\begin{figure}[ht]
    \centering
    \includegraphics[width=0.5\textwidth]{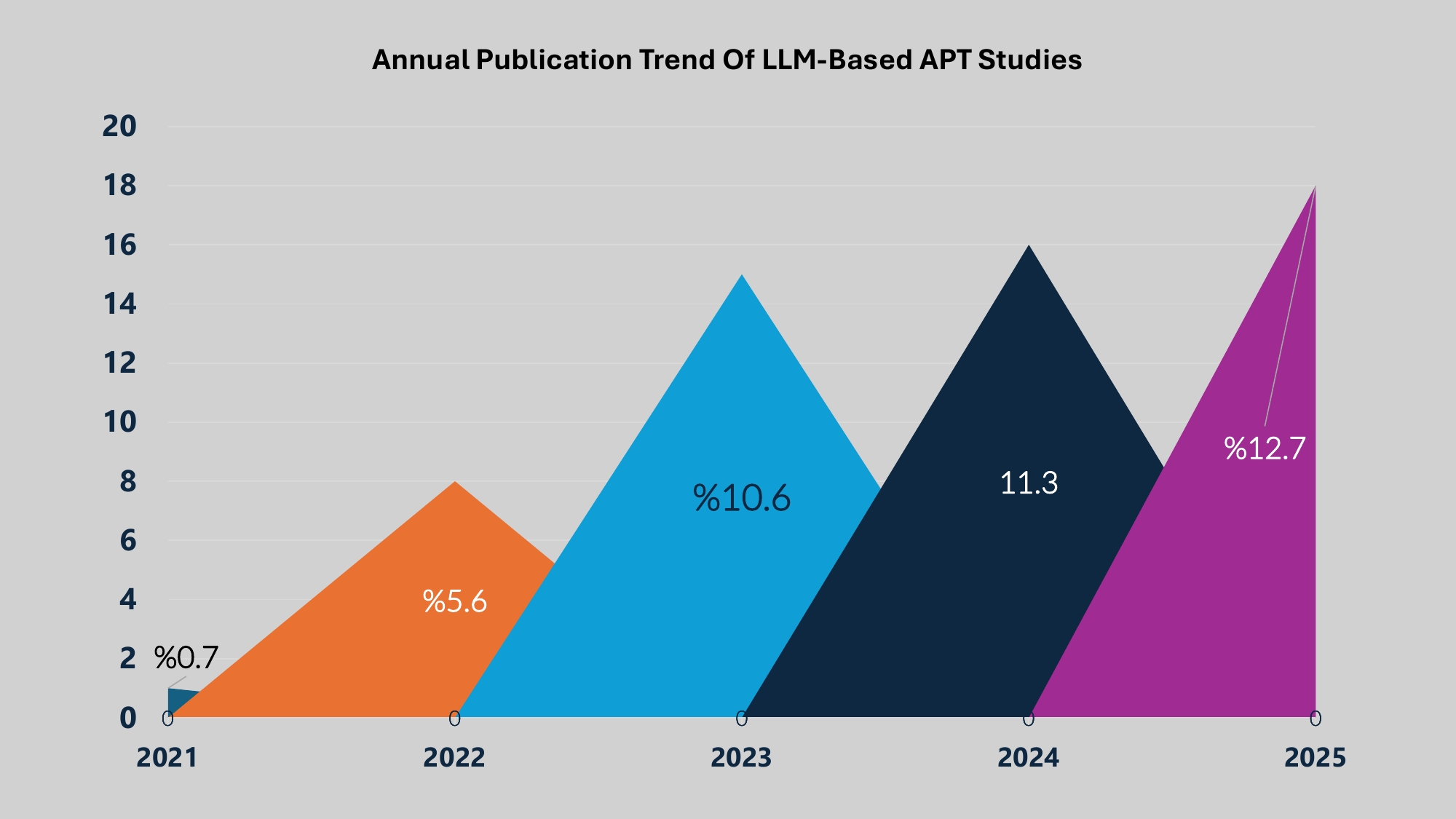}
    \caption{The Annual Trend showing the Percentage of LLM-based APT Studies published each Year}
    \label{fig:annual_publication_trend}
\end{figure}

\textbf{Year}: The last column shares the publication dates of the reviewed studies and evaluates the increase in LLM-focused APT papers as we move towards the 6G wireless networks era. Figure \ref{fig:annual_publication_trend} shows the change in LLM-focused APT papers by year.

\section{OPEN CHALLENGES AND FUTURE DIRECTIONS} \label{sec:challengesandfuture}

\begin{figure}[ht]
    \centering
    \includegraphics[width=0.5\textwidth]{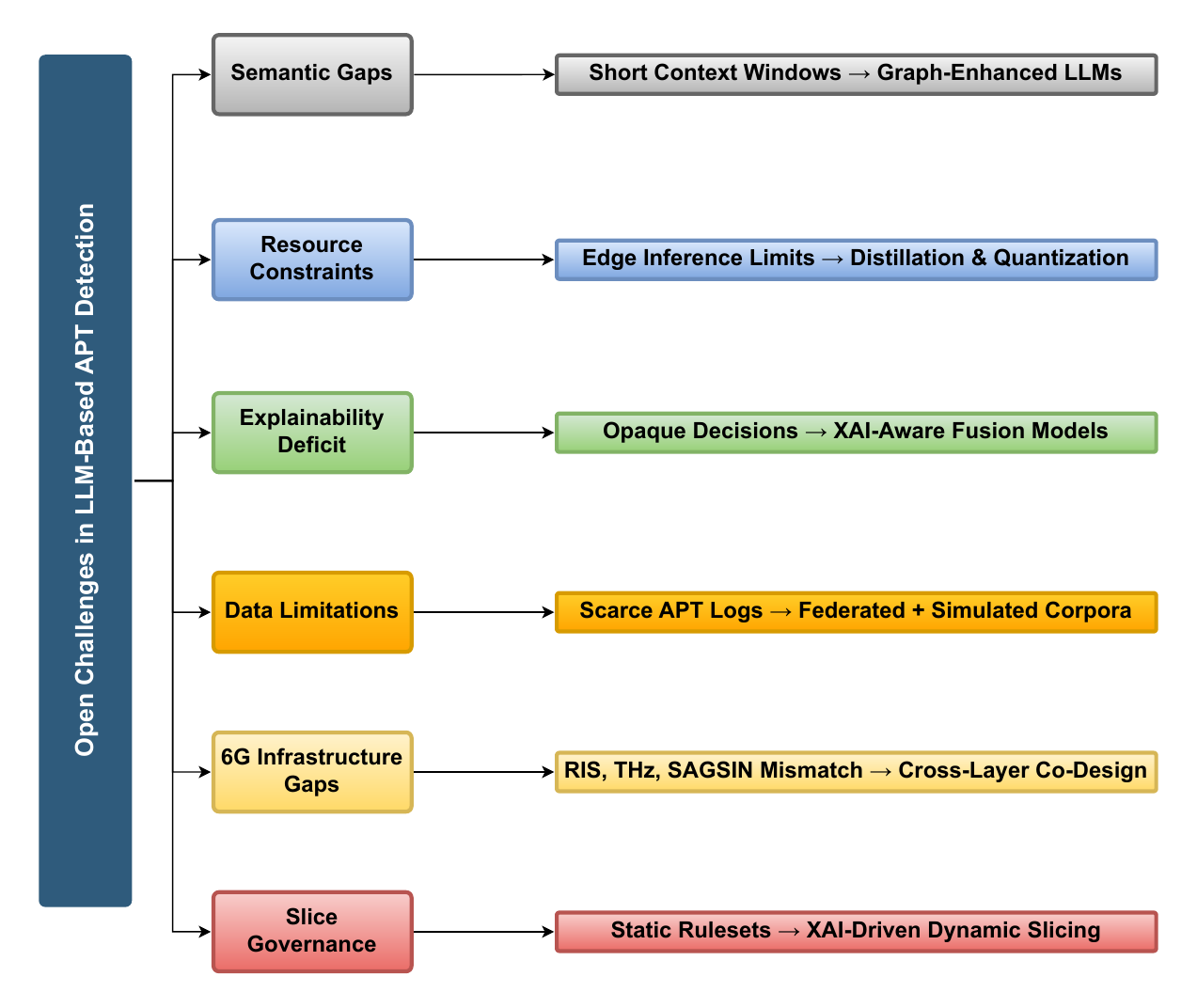}
    \caption{Taxonomy of Open Challenges in LLM-Based APT Detection}
    \label{fig:taxonomy_open_challenges}
\end{figure}

As the use of 6G networks and LLM deployments in 6G becomes widespread, many research gaps and new open challenges to be solved will emerge for researchers. These challenges include architecture and security issues, and we discuss the open challenges, a taxonomy of which is given in Figure \ref{fig:taxonomy_open_challenges}, in this section.

\textbf{Semantic-Aware Reasoning and Limited Contextual Memory}: LLMs are promising for APT detection in 6G networks with their high performance in understanding causal relationships and threat contexts using data such as system logs and audit trails \cite{oleiwi2023meta}. However, LLM models have limited performance in long-term and fragmented event sequences because their architectures offer limited window sizes and context management. Therefore, it makes detection difficult in multi-stage APTs with long processes such as infiltration and reconnaissance.

\textit{Future Directions}: Future researchers can overcome these limitations by focusing on the integration of memory modules and hierarchical memory structures. These structures make it easier for the model to learn long-term correlations between events. An example of this is the modeling of the relationship between system input and data exfiltration behavior to understand the holistic behavior of an attack. In addition, models that can provide transformer-graph synergy, such as GNNs, can effectively establish topological or temporal relationships between events \cite{15}. Thus, event traces can be modeled over a graph structure, enabling LLM models to learn event dependencies in a scalable manner.

\textbf{Real-Time Processing Under Edge Constraints}: 6G lines provide great advantages for time-constrained scenarios such as autonomous vehicles by offering high speed and low latency \cite{sergiou2020complex}. In particular, since edge devices in their heterogeneous structure bring the processing power closer to the data source, it will enable real-time data processing with low latency and low bandwidth usage \cite{yang2022security}. However, since LLM models require high processing power and most edge devices have low capacity and low processing power, this poses a serious challenge.

\textit{Future Directions}: Future researchers can work on threat-aware and edge-adaptive LLM models to overcome this challenge. For this, some techniques such as knowledge distillation, quantization, and edge-aware fine-tuning come to the fore. These techniques are explained in detail in section 5.3.

\textbf{Lack of Grounded Explainability}: Although LLM-based models have great potential in cybersecurity, such as APT detection, most models can cause serious security vulnerabilities in mission-critical tasks due to their black-box nature \cite{12}. For this reason, it is necessary to understand the inputs and probabilities that the model uses when making this decision. However, there is no system that shows the necessary causal traceability and root cause reasoning information in LLM models to make this understanding. This causes gaps in forensic analysis, such as explaining attacks and auditing sources.

\textit{Future Directions}: Future researchers can design more transparent systems by examining network slicing and decision-making processes for LLM models with new Explainable Artificial Intelligence (XAI) frameworks. More transparent information can be obtained with techniques that explain the training phase of models and the output phase of models, especially pre-hoc XAI and post-hoc XAI.

\textbf{Scarcity of Fine-Grained LLM Training Data}: Data quality has a major impact on the predictive performance of LLM models, and the volume and variety of APT-related data used in the existing literature are limited in terms of real-world representation \cite{ghiasvand2024cicapt}. Studies (see section 5.5) show that most studies rely on synthetic audit logs or CTIs with limited content. This limits the generalizability of LLM models and their ability to detect threats in different environments.

\textit{Future Directions}: To overcome this data limitation issue, steps can be taken such as collaboration between researchers and organizations (public-private), development of benchmark datasets, and modeling of attack progression scenarios.

\textbf{Integration with Emerging 6G Technologies}: In addition to high data rates in 6G networks, new generation technologies such as intelligent reflective surfaces (IRS) and terahertz (THz) band communication will also provide more dynamic and uninterrupted communication opportunities \cite{16}. However, this also brings some challenges such as synchronization, spectrum sharing, and secure orchestration. Adaptation of current LLM-based APT systems to such complex and multi-layered environments requires great attention.

\textit{Future Directions}: To overcome these challenges, researchers can develop multi-layered security protocols. In this way, 6G networks will gain threat perception and response cycle capability in ultra-dynamic and variable environments.

\textbf{Underexplored Role of Network Slicing and XAI Fusion}: With the network slicing feature, 6G networks can run mission-critical scenarios such as autonomous vehicle communication and industrial control on dedicated and isolated resources \cite{12}. However, incorrect resource allocations and unexpected load shifts that may occur during slicing operations can cause serious security problems \cite{abdulqadder2022sliceblock}. An example is the autonomous vehicle experiencing signal delays that are beyond the delay tolerance due to network slicing.

Although XAI techniques provide dynamic adaptation capabilities in network slicing, the use of these capabilities in real-time environments is still limited \cite{12}. Since most of the research is theory or simulation-based, its use with real-world data from SDN infrastructures needs to be investigated.

\textit{Future Directions}: To overcome these limitations, researchers can develop slice-aware and network state-oriented LLM models, and these models should be able to dynamically adjust network slice configurations and allocation policies by continuously monitoring real-time data. In addition, XAI techniques can be integrated with the obtained decisions to provide traceable and reliable information for network operators.

\section{CONCLUSIONS} \label{sec:conclusions}

This paper presents a comprehensive systematic review and taxonomy, the first of its kind, for LLM-based Advanced Persistent Threat (APT) detection in 6G networks. Findings from 142 recent papers examine the interaction between the capabilities (semantics) of LLM models and the challenges (architecture, privacy, etc.) of 6G environments. We aim to provide new insights for future research by presenting a taxonomy covering input types, model techniques, deployment settings, and threat lifecycle stages. Although LLM has great potential in APT attack detection, it also has limitations such as limited context memory, opaque decision processes, and real-time inference at the edge. In addition, reproducibility and dataset generalizability stand out as important obstacles for research in this area. Based on the findings, we call for joint efforts in the following research areas:

\begin{itemize}

\item  Designing lightweight, unified LLMs for edge devices in 6G networks,

\item  Investigating new XAI-driven decision monitoring mechanisms to increase transparency of LLMs,

\item  Enriching datasets used for APT detection using fine-grained, multimodal, and real-world data,

\item Integrating LLMs with slicing-aware orchestration systems in 6G for dynamic demands on 6G links.

\end{itemize}

\section{Appendix A: Research Selection Criteria and Article Overview}

After a comprehensive literature search and systematic analysis (Kitchenham's Systematic Literature Review (SLR) approach and Petersen's Systematic Mapping Study (SMS) ), we used the form in Table \ref{table:llm_apt_eval} to select the most relevant and high-quality articles from 142 obtained articles. The questions in this form were used to select the articles focusing on LLM-based APT detection solutions among the publications addressing the Advanced Persistent Threat (APT) problem in the context of 6G wireless networks. Also, the lists of all these articles are provided in Table \ref{tab:rq1_llm_fragmented_provenance}, \ref{tab:rq2_llm_fragmented_provenance}, \ref{tab:llm_edge_compression_6g}, \ref{tab:llm_apt_datasets_survey}.

\begin{table}[ht]
\centering
\caption{Research Evaluation Questions for LLM-Based APT Detection Studies in 6G Networks}
\label{table:llm_apt_eval}
\resizebox{\textwidth}{!}{%
\begin{tabular}{|p{12cm}|c|c|}
\hline
\textbf{Question}                                                                                  & \textbf{YES} & \textbf{NO} \\ \hline
Does the reviewed article detect APTs in 6G networks?                                              &              &             \\ \hline
Does the reviewed article include an LLM-based solution method?                                   &              &             \\ \hline
Does the reviewed article provide an approach or methodology that includes LLMs for APT detection? &              &             \\ \hline
Does the proposed method work under 6G-related constraints (such as edge resource constraints)?    &              &             \\ \hline
Is an optimization method suggested? If yes, which method (distillation etc.)?                    &              &             \\ \hline
Are there datasets or simulated environments for LLM-based APT detection?                         &              &             \\ \hline
Is the source code and/or dataset shared for reproducibility?                                     &              &             \\ \hline
What is the publication location and year?                                                        &              &             \\ \hline
\end{tabular}%
}
\end{table}

\begin{table}[ht]
\centering
\caption{Overview of 58 Studies Related to RQ1}
\label{tab:rq1_llm_fragmented_provenance}
\resizebox{\textwidth}{!}{%
\begin{tabular}{|c|c|c|c|c|c|c|}
\hline
\textbf{No} & \textbf{Paper}                         & \textbf{Code Availability} & \textbf{Dataset}             & \textbf{Eval. Protocol}  & \textbf{Venue/Platform}    & \textbf{Year} \\ \hline
\textbf{1}  & Koenders (1.pdf)                       & NO                         & CVE/ATT\&CK                  & YES                      & Erasmus MSc                & 2024          \\ \hline
\textbf{2}  & Zuo et al. (2.pdf)                     & NO                         & audit data implied           & YES                      & Sandia/UCO                 & 2025          \\ \hline
\textbf{3}  & Cheng et al. (3.pdf)                   & GitHub                     & benchmark                    & OMNISEC eval             & arXiv                      & 2025          \\ \hline
\textbf{4}  & Xu et al. (4.pdf)                      & NO                         & public datasets              & survey w/ metrics        & ACM JACM                   & 2025          \\ \hline
\textbf{5}  & Ali et al. (5.pdf)                     & NO                         & NO                           & NO                       & SHIFRA Journal             & 2025          \\ \hline
\textbf{6}  & Zhang et al. (6.pdf)                   & NO                         & NO                           & NO                       & TechRxiv (preprint)        & 2025          \\ \hline
\textbf{7}  & Jeon et al. (7.pdf)                    & NO                         & NO                           & RAG-based graph eval     & Conf. Paper (Korea)        & 2025          \\ \hline
\textbf{8}  & Blänsdorf (8.pdf)                      & NO                         & CTI list                     & manual labeling          & MSc Thesis (Chalmers)      & 2024          \\ \hline
\textbf{9}  & Yin et al. (9.pdf)                     & NO                         & NO                           & NO                       & arXiv                      & 2025          \\ \hline
\textbf{10} & Antar (10.pdf)                         & ECHO env.                  & NO                           & PromptPilot eval         & MSc Thesis (Queen’s)       & 2025          \\ \hline
\textbf{11} & Xu et al. (11.pdf)                     & NO                         & CTI reports                  & IntelEX framework        & arXiv                      & 2025          \\ \hline
\textbf{12} & Chen et al. (12.pdf)                   & AECR pipeline              & NO                           & F1, precision, recall    & Elsevier CoSE              & 2025          \\ \hline
\textbf{13} & Tan et al. (13.pdf)                    & NO                         & NO                           & temporal graph eval      & Conf. (Glasgow)            & 2025          \\ \hline
\textbf{14} & Tan et al. (14.pdf / 16.pdf)           & NO                         & NO                           & survey + taxonomy        & IEEE IoT Journal           & 2025          \\ \hline
\textbf{15} & Purba (15.pdf)                         & NO                         & NO                           & Kibana-query generation  & PhD Diss. (UNC Charlotte)  & 2025          \\ \hline
\textbf{16} & Wang et al. (17.pdf)                   & pending                    & APT dataset >1k              & AURORA system            & arXiv                      & 2025          \\ \hline
\textbf{17} & Kavousi (18.pdf)                       & NO                         & NO                           & semantic security eval   & PhD Diss. (Northwestern)   & 2025          \\ \hline
\textbf{18} & Zhang \& Tenney (19.pdf)               & NO                         & NO                           & NO (survey only)         & OJBM                       & 2024          \\ \hline
\textbf{19} & Daniel et al. (20.pdf)                 & NO                         & Snort rules                  & LLM vs. ML eval          & MDPI BDCC                  & 2025          \\ \hline
\textbf{20} & Ahmed (21.pdf)                         & prototype impl.            & DARPA OpTC                   & distributed eval         & PhD Diss. (UNC Charlotte)  & 2024          \\ \hline
\textbf{21} & Du et al. (22.pdf)                     & NO                         & benchmark used               & MAD-LLM eval             & IEEE ISPA                  & 2024          \\ \hline
\textbf{22} & Mezzi et al. (23.pdf)                  & Eval framework             & 350 CTI reports              & calibration, consistency & arXiv                      & 2025          \\ \hline
\textbf{23} & Suomalainen et al. (24.pdf)            & NO                         & Cyber Ops Tracker            & LLM for CTI metrics      & TechRxiv                   & 2025          \\ \hline
\textbf{24} & Alturkistani (25.pdf)                  & NO                         & NO                           & systematic SLR analysis  & Research Square (SLR)      & 2024          \\ \hline
\textbf{25} & Sultana et al. (26.pdf)                & NO                         & NO                           & LLM eval. framework      & IEEE CNS Workshop          & 2023          \\ \hline
\textbf{26} & Cui et al. (27.pdf)                    & NO                         & NO                           & LLM risk tax. + eval     & Tsinghua Lab + Ant Group   & 2024          \\ \hline
\textbf{27} & Li et al. (28.pdf)                     & NO                         & NO                           & agentic eval + 5G        & NYU Tech Report (arXiv)    & 2025          \\ \hline
\textbf{28} & Daniel et al. (29.pdf)                 & Snort parser               & 973 rules dataset            & LLM vs ML accuracy       & arXiv                      & 2024          \\ \hline
\textbf{29} & Mitra et al. (30.pdf)                  & NO                         & NO                           & LocalIntel eval          & arXiv                      & 2025          \\ \hline
\textbf{30} & Wang et al. (31.pdf) – MultiKG         & MultiKG repo               & real CTI + logs              & cross-source KG eval     & arXiv                      & 2024          \\ \hline
\textbf{31} & Yao (32.pdf)                           & 5GSecRec impl.             & 5G+Kube alerts               & QA + correlation         & MSc Thesis (Concordia)     & 2024          \\ \hline
\textbf{32} & Mahboubi et al. (33.pdf)               & NO                         & open ontologies              & survey + ML eval         & Elsevier JNCA              & 2024          \\ \hline
\textbf{33} & Sammouri (34.pdf) – CAPEC model        & CAPEC model                & CAPEC taxonomy               & expert evaluation        & MSc Thesis (Miami Univ.)   & 2025          \\ \hline
\textbf{34} & Sewak et al. (35.pdf)                  & NO                         & NO                           & threat graph eval        & CIKM Workshop              & 2023          \\ \hline
\textbf{35} & Kasri et al. (36.pdf)                  & NO                         & NO                           & review + cases           & MDPI Computation           & 2025          \\ \hline
\textbf{36} & Hasanov et al. (37.pdf)                & NO                         & NO                           & SLR criteria             & IEEE Access                & 2024          \\ \hline
\textbf{37} & Keltek et al. (38.pdf) – LSAST         & LSAST prototype            & HackerOne dump               & LLM vs SAST              & arXiv                      & 2024          \\ \hline
\textbf{38} & Jawad (39.pdf)                         & NO                         & NO                           & PhD eval phases          & PhD Plan (Spain)           & 2025          \\ \hline
\textbf{39} & Würsch et al. (40.pdf)                 & NO                         & arXiv NLP corpus             & NER/ER comparison        & arXiv                      & 2023          \\ \hline
\textbf{40} & Diakhame et al. (41.pdf) – MCM-Llama   & LLM pipeline               & security events              & LLM vs NER/Sim           & ICECET (IEEE Conf.)        & 2024          \\ \hline
\textbf{41} & Chen et al. (42.pdf) – LLM Survey      & NO                         & NO                           & systematic stages review & Elsevier CoSE              & 2024          \\ \hline
\textbf{42} & Rahman (43.pdf) – Incident Reconstruct & local LLM sys.             & PCAP, ChromaDB               & LLM eval on NTA          & MSc Thesis (Turku)         & 2024          \\ \hline
\textbf{43} & Ji et al. (44.pdf) – SEVENLLM          & GitHub                     & 28-task benchmark            & multi-task eval          & arXiv                      & 2024          \\ \hline
\textbf{44} & Taghavi et al. (45.pdf) – LLM4Vuln     & NO                         & NO                           & survey + workflow eval   & ResearchSquare (preprint)  & 2024          \\ \hline
\textbf{45} & Sood et al. (46.pdf) – Hallucination   & NO                         & NO                           & taxonomy + mitigation    & Elsevier CEE               & 2025          \\ \hline
\textbf{46} & Zhang et al. (47.pdf) – GENTTP         & Tool released              & PyPI Malware + GT            & Zero-shot eval, chatbot  & arXiv                      & 2024          \\ \hline
\textbf{47} & Fayyazi et al. (48.pdf) – TTP-LLM      & GitHub                     & NO                           & RAG vs SFT benchmark     & IEEE ACSAC Workshops       & 2024          \\ \hline
\textbf{48} & Zhang et al. (49.pdf) – UniTTP         & NO                         & Internal + public            & F1, task-level           & IEEE TrustCom              & 2024          \\ \hline
\textbf{49} & Arikkata et al. (50.pdf) – DroidTTP    & RAG pipeline               & Android TTP Dataset          & Jaccard, Hamming Loss    & arXiv                      & 2025          \\ \hline
\textbf{50} & Li et al. (51.pdf) – AnomalyGen        & LogSynth code              & synthetic logs               & F1 gain eval             & Conf. Paper (preprint)     & 2025          \\ \hline
\textbf{51} & Shan et al. (52.pdf) – LogConfigLocal  & tool available             & Hadoop logs                  & root-cause accuracy      & ACM ISSTA                  & 2024          \\ \hline
\textbf{52} & Huang et al. (53.pdf) – LUNAR          & GitHub                     & LCU logs                     & log parsing eval         & ASE Conf. (arXiv preprint) & 2024          \\ \hline
\textbf{53} & He et al. (54.pdf) – LLMeLog           & fine-tuned BERT            & 3 public logs                & F1 > 99\%                & IEEE ISSRE                 & 2024          \\ \hline
\textbf{54} & Wang et al. (55.pdf) – LM Agents       & NO                         & NO                           & multi-agent eval         & arXiv                      & 2025          \\ \hline
\textbf{55} & Balasubramanian et al. (56.pdf)        & chatbot code               & open logs used               & GPT-3 vs others eval     & IEEE BigData               & 2023          \\ \hline
\textbf{56} & Gandhi et al. (57.pdf) – SHIELD        & NO                         & NO (custom logs only)        & precision/recall         & arXiv                      & 2025          \\ \hline
\textbf{57} & Benabderrahmane et al. (58.pdf)        & NO                         & DARPA TC                     & AE/VAE eval              & arXiv                      & 2025          \\ \hline
\textbf{58} & Ferrag et al. (59.pdf) – LLM Survey    & NO                         & NO                           & survey w/ benchmarks     & SSRN (preprint)            & 2025          \\ \hline
\end{tabular}%
}
\end{table}

\begin{table}[ht]
\centering
\caption{Overview of 28 Studies Related to RQ2}
\label{tab:rq2_llm_fragmented_provenance}
\resizebox{\textwidth}{!}{%
\begin{tabular}{|c|c|c|c|c|c|c|}
\hline
\textbf{No} & \textbf{Paper}                         & \textbf{Code Availability} & \textbf{Dataset}             & \textbf{Eval. Protocol}  & \textbf{Venue/Platform}    & \textbf{Year} \\ \hline
1  & Albshaier et al. (Ä.pdf)                  & NO                          & Not specified (SLR only)                          & YES                                     & Electronics (MDPI)                   & 2025          \\ \hline
2  & Hameed et al. (A.pdf)                     & NO                          & IoT-based (SMPC, DP, HE)                          & Preprint (SSRN)                         & SSRN / Nuclear Phys. B               & 2025          \\ \hline
3  & Al-Kadhimi et al. (a1.pdf)                & NO                          & Mobile APT data (1351 reviewed)                   & YES (SLR + Framework)                  & Applied Sciences (MDPI)              & 2023          \\ \hline
4  & Le \& Shetty (B.pdf)                      & NO                          & 5G-based IoT                                      & Conceptual (no benchmarks)             & Ad Hoc Networks (Elsevier)           & 2021          \\ \hline
5  & Alkaeed et al. (C.pdf)                    & NO                          & AI-XR / Metaverse data                            & YES (Survey)                           & J. of Network and Comp. Apps         & 2024          \\ \hline
6  & Xu et al. (Ç.pdf)                         & NO                          & Multiple (127 reviewed)                           & YES (Survey)                           & LLM4Security Survey                  & 2024          \\ \hline
7  & Zhang et al. (D.pdf)                      & YES (planned)               & 4 RS models (LLM-based)                           & Experiments on 4 models                & arXiv                                 & 2024          \\ \hline
8  & Guo et al. (E.pdf)                        & YES                         & Various LLMs (LLaMA, GPT)                         & Extensive experiments                  & ICML (PMLR)                           & 2024          \\ \hline
9  & Cao et al. (F.pdf)                        & NO                          & Fine-tuned LLMs (backdoor)                        & Experiments (safety test bypass)       & arXiv                                 & 2024          \\ \hline
10 & Aguilera-Martínez \& Berzal (G.pdf)       & NO                          & Training + Inference threats                      & YES (Survey)                           & arXiv                                 & 2025          \\ \hline
11 & Lanka et al. (O.pdf)                      & NO                          & Honeypot + UEBA data                              & YES (LLM-based analysis)              & Electronics (MDPI)                   & 2024          \\ \hline
12 & Zhang et al. (H.pdf)                      & NO                          & Poisoned RAG documents                            & YES                                     & FSE Companion (ACM)                  & 2024          \\ \hline
13 & Rahman \& Hossain (I.pdf)                 & NO                          & IIoT logs in 6G                                   & YES (SDS + DL)                         & IEEE Wireless Comm.                  & 2022          \\ \hline
14 & Wang et al. (ı.pdf)                       & NO                          & MEC + AI security logs                            & YES (ETSI-based survey)               & IEEE IoT Journal                     & 2023          \\ \hline
15 & Alevizos et al. (İ.pdf)                   & NO                          & Blockchain-based IDS in VSNs                      & YES (Throttling eval)                 & Sensors (MDPI)                        & 2023          \\ \hline
16 & Nahar et al. (J.pdf)                      & NO                          & ZTA in 6G                                         & YES (Use case studies)                & IEEE Access                          & 2024          \\ \hline
17 & Je et al. (K.pdf)                         & NO                          & Open 6G + AI systems                              & YES (Threat mapping)                  & IEEE Comm. Standards Mag.            & 2021          \\ \hline
18 & Xu et al. (L.pdf)                         & YES                         & APT Traffic (Anyrun2024)                          & YES                                     & CAS / UCAS                           & 2024          \\ \hline
19 & Du et al. (M.pdf)                         & NO                          & Multi-source alerts                               & YES                                     & IEEE ISPA 2024                       & 2024          \\ \hline
20 & Liu et al. (N.pdf)                        & YES                         & Contextual demos for agents                       & YES (Backdoor trigger eval)           & IEEE TIFS                            & 2025          \\ \hline
21 & Hassanin \& Moustafa (Z.pdf)              & NO                          & Survey on LLM for cyber defense                   & YES                                     & arXiv                                 & 2024          \\ \hline
22 & Mao et al. (Ö.pdf)                        & NO                          & Edge computing / cache / intelligence             & YES                                     & IEEE COMST                           & 2023          \\ \hline
23 & Yang et al. (P.pdf)                       & NO                          & 6G Security Protocols                             & YES                                     & arXiv / ACM                          & 2024          \\ \hline
24 & Hadi et al. (R.pdf)                       & YES                         & UAVIDS / NF-UQ / 5G-NIDD                          & YES                                     & Expert Systems w/ Applications       & 2024          \\ \hline
25 & Chen et al. (Ş.pdf)                       & NO                          & Various threat logs + LLMs                        & YES                                     & Computers \& Security                 & 2024          \\ \hline
26 & Sun et al. (T.pdf)                        & YES                         & DL model evasions (semantic traffic)              & YES                                     & WWW 2025                             & 2025          \\ \hline
27 & Diao et al. (V.pdf)                       & NO                          & DoH tunnel traffic                                & YES (Feature fusion, Recall: 0.9995) & ACM CCS Poster                       & 2024          \\ \hline
28 & Sun et al. (Y.pdf)                        & YES                         & Adversarial LLM traffic (6 datasets)              & YES (RL + Payload tuning)            & WWW 2025                             & 2025    \\ \hline
\end{tabular}%
}
\end{table}

\begin{table}[ht]
\centering
\caption{Overview of 26 Studies Related to RQ3}
\label{tab:llm_edge_compression_6g}
\resizebox{\textwidth}{!}{%
\begin{tabular}{|c|c|c|c|c|c|c|}
\hline
\textbf{No} & \textbf{Paper}                     & \textbf{Code Availability} & \textbf{Dataset}                           & \textbf{Eval. Protocol}                         & \textbf{Venue/Platform}              & \textbf{Year} \\ \hline
1  & Liu et al. (a1.pdf)              & NO                          & Not specified                              & Review (model compression only)                 & Frontiers in Robotics and AI         & 2025          \\ \hline
2  & Friha et al. (a2.pdf)            & NO                          & Not specified                              & Comprehensive survey                           & IEEE OJCOMS                          & 2024          \\ \hline
3  & Zhang et al. (a3.pdf)            & NO                          & Llama2                                     & Real testbed with Llama2 models                & IEEE IoT Journal                      & 2025          \\ \hline
4  & Cai et al. (a4.pdf)              & NO                          & LLaMA, ChatGLM                             & Edge-LLM framework eval                        & Conference (not stated)               & 2025          \\ \hline
5  & Qu et al. (a10.pdf)             & NO                          & Not specified                              & Survey on MEI for LLMs                         & IEEE COMST (accepted)                 & 2025          \\ \hline
6  & Picano et al. (a6.pdf)           & NO                          & Testbed for autonomous driving             & Matching-based optimization                    & IEEE OJCOMS                          & 2025          \\ \hline
7  & Ray \& Pradhan (a9.pdf)         & NO                          & IoT edge, quantized LLMs                   & LLMEdge Framework Demo                         & Not stated                            & 2025          \\ \hline
8  & Kim et al. (a20.pdf)            & NO                          & Not specified                              & Systematic review on compression \& tuning     & ACM Computing Surveys                 & 2025          \\ \hline
9  & Zhang et al. (a11.pdf)          & NO                          & Not specified                              & Quantization + batching (wireless constraint)  & IEEE TWC                              & 2025          \\ \hline
10 & Wei et al. (a12.pdf)            & YES (T-MAC GitHub)          & LLaMA, BitNet                              & LUT-based low-bit benchmarking                 & EuroSys 2025                         & 2025          \\ \hline
11 & Semerikov et al. (a13.pdf)      & NO                          & LLM-edge cases, EdgeLLM, EdgeShard         & Comprehensive Survey                           & CEUR Workshop Proceedings             & 2025          \\ \hline
12 & Dhar et al. (a14.pdf)           & NO                          & LLaMA-2 7B INT4                            & Empirical edge inference eval                  & ACMSE 2024                            & 2024          \\ \hline
13 & Zheng et al. (a15.pdf)          & NO                          & Meta LLaMA, DeepSeek                       & Lifecycle review + hardware co-design          & ACM Computing Surveys                 & 2025          \\ \hline
14 & Qin et al. (a16.pdf)            & NO                          & LaMP datasets                              & Empirical guidelines + compression comparison  & arXiv (Preprint)                      & 2025          \\ \hline
15 & Yang et al. (a17.pdf)           & NO                          & Not specified                              & PerLLM edge-cloud scheduling                   & arXiv (Preprint)                      & 2024          \\ \hline
16 & Liu et al. (a18.pdf)            & NO                          & Not specified                              & Fractional programming optimization            & MOBIHOC '24                           & 2024          \\ \hline
17 & Zhu et al. (a19.pdf)            & NO                          & Not specified                              & Compression taxonomy (quant., prune, KD)       & TACL                                   & 2024          \\ \hline
18 & Zhao et al. (a29.pdf)           & NO                          & Not specified                              & Edge-terminal token decoding optimization      & IEEE (Wireless)                       & 2024          \\ \hline
19 & Wang et al. (a21.pdf)           & NO                          & Not specified                              & Inference taxonomy (compression focus)         & IEEE (Preprint)                       & 2024          \\ \hline
20 & Wang et al. (a22.pdf)           & NO                          & Quantized OPT-1.3B                         & Compression-aware quantization + pruning       & arXiv (Preprint)                      & 2025          \\ \hline
21 & Xu et al. (a23.pdf)             & YES (GitHub)                & Multiple LLMs                              & Multi-dimensional safety evaluation            & arXiv (Preprint)                      & 2024          \\ \hline
22 & Liu et al. (a24.pdf)            & NO                          & GPTQ, SmoothQuant                          & Survey on efficient training/inference         & arXiv (Preprint)                      & 2025          \\ \hline
23 & Wen et al. (a25.pdf)            & NO                          & Custom telemetry logs                      & LLM-based anomaly detection in 6G              & HOTNETS '24                           & 2024          \\ \hline
24 & Khowaja et al. (a26.pdf)        & NO                          & Named Entity Recognition (NER)             & Membership inference on ZSM fine-tuning        & IEEE (Preprint)                       & 2024          \\ \hline
25 & Qin et al. (a27.pdf)            & NO                          & Multiple (SAGIN datasets)                  & CoT-based security for 6G SAGIN                & IEEE (Preprint)                       & 2025          \\ \hline
26 & Qu et al. (a28.pdf)             & NO                          & MEI4LLM (LLMs + Edge)                      & Survey + MEI framework                         & IEEE COMST (Accepted)                 & 2025          \\ \hline
\end{tabular}%
}
\end{table}

\begin{table}[ht]
\centering
\caption{Overview of 30 Studies Related to RQ4}
\label{tab:llm_apt_datasets_survey}
\resizebox{\textwidth}{!}{%
\begin{tabular}{|c|l|c|c|l|l|c|}
\hline
\textbf{No} & \textbf{Paper}                          & \textbf{Code Availability} & \textbf{Dataset}                           & \textbf{Eval. Protocol}                          & \textbf{Venue/Platform}             & \textbf{Year} \\ \hline
1  & Abu Talib et al. (2.pdf)              & NO                          & YES                                       & YES (Systematic Review)                         & Computers \& Security               & 2022          \\ \hline
2  & Abu Talib et al. (23.pdf)             & NO                          & NO                                        & YES (Review on APT Beaconing)                  & Computers \& Security               & 2022          \\ \hline
3  & Al-Aamri et al. (32.pdf)              & NO                          & YES (Custom logs + flow)                 & YES (CDT model + IDS)                           & Sustainability (MDPI)              & 2023          \\ \hline
4  & Do Xuan \& Nam (3.pdf)                & NO                          & NO                                        & YES (Domain monitoring)                        & Procedia Computer Science           & 2019          \\ \hline
5  & Do Xuan \& Nguyen (30.pdf)            & NO                          & YES (APT IP traffic)                     & YES (BiLSTM + Attention + DGCNN)               & Scientific Reports                 & 2024          \\ \hline
6  & Do Xuan et al. (4.pdf)                & NO                          & YES (Reconstructed flows)               & YES (BiLSTM-GCN based)                         & J. Intelligent \& Fuzzy Syst.      & 2020          \\ \hline
7  & El Alami \& Rawat (16.pdf)            & NO                          & YES (TON-IoT)                            & YES (GAN, LSTM, AE eval)                       & —                                   & 2024          \\ \hline
8  & Ferrag et al. (11.pdf)                & NO                          & YES (FL/Edge datasets)                   & YES (Comprehensive Survey)                     & IEEE Com. Surveys \& Tutorials     & 2023          \\ \hline
9  & Ferrag et al. (27.pdf)                & NO                          & YES (42 models \& datasets)             & YES (Taxonomy + LLM Eval)                      & SSRN (preprint)                    & 2024          \\ \hline
10 & Ghiasvand et al. (25.pdf)             & YES                         & YES (CICAPT-IIoT Dataset)               & YES (Multi-phase APT dataset)                 & arXiv                               & 2024          \\ \hline
11 & Gupta et al. (15.pdf)                 & YES (PyTorch model)         & YES (KDDCup)                             & YES (DoS, Probe, Sybil attacks)               & —                                   & 2024          \\ \hline
12 & Huang et al. (24.pdf)                 & YES (SAGA)                  & YES (Synthetic Logs)                    & YES (Technique-hunting, APT lifecycle)        & —                                   & 2024          \\ \hline
13 & Motlagh et al. (19.pdf)               & NO                          & NO                                        & YES (Offensive \& defensive use)              & arXiv                               & 2024          \\ \hline
14 & Myneni et al. (20.pdf)                & NO                          & YES (Unraveled dataset)                 & YES (Semi-synthetic APT eval)                 & Computer Networks                  & 2023          \\ \hline
15 & Neuschmied et al. (31.pdf)            & NO                          & YES (Contagio + CICIDS2017)             & YES (AE + VAE models)                         & Applied Sciences (MDPI)            & 2022          \\ \hline
16 & Nezhadsistani \& Stiller (8.pdf)      & NO                          & YES                                      & YES (Survey, Challenges, Metrics)             & IEEE 6GNet                          & 2024          \\ \hline
17 & Nguyen et al. (17.pdf)                & NO                          & NO                                        & YES (Threat taxonomy, LLMSecOps)              & arXiv                               & 2024          \\ \hline
18 & Nguyen et al. (5.pdf)                 & NO                          & YES                                      & YES (MIG: MLP + Inference + GCN)              & J. Intelligent \& Fuzzy Syst.      & 2023          \\ \hline
19 & Oleiwi et al. (28.pdf)                & NO                          & YES (NSL-KDD, CIC, etc.)                & YES (Stacked meta-model, classifiers)         & Electronics (MDPI)                 & 2023          \\ \hline
20 & Rajendran \& Vyas (6.pdf)             & YES                         & YES (Custom)                             & YES (Comparative Eval)                        & SoutheastCon                        & 2024          \\ \hline
21 & Saeed et al. (29.pdf)                 & NO                          & YES (Multiple APT datasets)             & YES (Hybrid EL, CFS-RF, Adaboost)             & Electronics (MDPI)                 & 2023          \\ \hline
22 & Shafee et al. (18.pdf)                & NO                          & YES (Twitter-CTI)                        & YES (Binary Class. + NER)                     & Expert Syst. with Applications     & 2025          \\ \hline
23 & Sharma \& Rani (14.pdf)               & NO                          & YES (RT-IoT)                             & YES (Stacked-Hybrid ML eval)                  & IEEE IoT Journal                    & 2024          \\ \hline
24 & Stojanović et al. (1.pdf)             & NO                          & YES                                      & YES (Review of 20+ datasets)                  & Computers \& Security               & 2020          \\ \hline
25 & Unnamed (22.pdf)                      & NO                          & YES (Semi-synthetic network traffic)    & YES (ML early detection comparison)           & LLM-Aided (Unspecified)            & 2024          \\ \hline
26 & Viswanathan et al. (21.pdf)           & NO                          & YES (Partially synthetic MRI)           & YES (ML vs full/partial synthetic eval)       & BioMed (MRI Imaging)               & 2024          \\ \hline
27 & Xiong et al. (13.pdf)                 & NO                          & YES (Windows-host logs)                 & YES (FSM + Real-time eval)                    & IEEE TDSC                           & 2022          \\ \hline
28 & Xylouris et al. (7.pdf)               & YES (XGBoost, CNN, etc.)    & YES (5G Testbed Dataset)                & YES (Real-time SHAP Eval)                     & IEEE TCE                            & 2024          \\ \hline
29 & Yamin et al. (10.pdf)                 & YES (CyExec - GPT)          & NO                                        & YES (Scenario Gen, RAG prompts)              & IEEE Access                         & 2024          \\ \hline
30 & Zhang et al. (26.pdf)                 & NO                          & YES                                      & YES (PADASYN + AdaBoost eval)                & —                                   & 2024          \\ \hline
\end{tabular}%
}
\end{table}

\section*{Acknowledgements}

This work was supported by Zayed University Research Office, Fund number 23153.

\bibliographystyle{unsrt}
\bibliography{Ref}

\end{document}